\documentclass[journal,10pt]{IEEEtran}

\IEEEoverridecommandlockouts
\usepackage{cite}
\usepackage{caption}
\usepackage{amsmath,amssymb,amsfonts}
\usepackage{algorithmic}
\usepackage{graphicx}
\usepackage{tabularx}
\usepackage{textcomp}
\usepackage{xcolor}
\usepackage{stfloats}
\usepackage{multirow}
\usepackage{amsmath}
\usepackage{booktabs}
\usepackage{array}
\def\BibTeX{{\rm B\kern-.05em{\sc i\kern-.025em b}\kern-.08em
		T\kern-.1667em\lower.7ex\hbox{E}\kern-.125emX}}
\begin{document}
\title{A 3D Non-Stationary Channel Model for 6G Wireless Systems Employing Intelligent Reflecting Surfaces with Practical Phase Shifts}
	
	\author{Yingzhuo Sun, Cheng-Xiang Wang, \textit{Fellow, IEEE}, Jie Huang, \textit{Member, IEEE}, and Jun Wang, \textit{Student Member, IEEE}
	\thanks{Y. Sun, C.-X. Wang (corresponding author), J. Huang, and J. Wang are with National Mobile Communications Research Laboratory, School of Information Science and Engineering, Southeast University, Nanjing, 210096, China, and also with the Purple Mountain Laboratories, Nanjing, 211111, China (email: {\{sunyingzhuo, chxwang, j\_huang, jun.wang\}@seu.edu.cn}). }
	\thanks{This work was supported by the National Key R\&D Program of China under Grant 2018YFB1801101, the National Natural Science Foundation of China (NSFC) under Grants 61960206006 and 61901109, the Frontiers Science Center for Mobile Information Communication and Security, the High Level Innovation and Entrepreneurial Research Team Program in Jiangsu, the High Level Innovation and Entrepreneurial Talent Introduction Program in Jiangsu, the Research Fund of National Mobile Communications Research Laboratory, Southeast University, under Grant 2020B01, the Fundamental Research Funds for the Central Universities under Grant 2242020R30001, the Huawei Cooperation Project, and the EU H2020 RISE TESTBED2 project under Grant 872172.}}
\maketitle
	
\begin{abstract}
	In this paper, a three-dimensional (3D) geometry based stochastic model (GBSM) for a massive multiple-input multiple-output (MIMO) communication system employing practical discrete intelligent reflecting surface (IRS) is proposed. The proposed channel model supports the scenario where both transceivers and environments move. The evolution of clusters in the space domain and the practical discrete phase shifts are considered in the channel model. The steering vector is set at the base station for the cooperation with IRS. Through studying statistical properties, the non-stationary properties are verified. We find that IRS plays a role in separating the whole channel and make the absolute value of time autocorrelation function (ACF) larger than the situation without employing IRS. Time ACF of the case using discrete phase shifts is also compared with the continuous case.
\end{abstract}
\section{Introduction}
	With the advent of a new information era, larger volume of data, higher transmission rate, and better quality of service are appealed by users \cite{You1},\cite{6Gchannel}. The fifth generation (5G) wireless communication has been commercially used around the world recently. Although the most appropriate applications matching with 5G have not been developed well yet. But the potential benefits that 5G will bring to the society and economics deserve looking forward to. The peak data rate of 5G is about twenty times as the fourth generation (4G) wireless communication. The connection density is around ten times as 4G. We can clearly see the amazing advance 5G achieves. But the cost of this large performance promotion is the large economic cost and the huge power cost on account of the large amount of antennas, denser base stations (BS), and higher transmitting power. At the same time, the communication system is more complicated than before. We need to overcome these disadvantages during developing the sixth generation (6G) wireless communication. The researchers have proposed some solutions to get it. Among all these new solutions, intelligent reflecting surface (IRS) draws much attention due to its unique advantages. IRS is a category of metamaterial which is made of sub-wavelength elements arranged in a specific order \cite{cui1}. It has some properties totally different from material in the nature, such as negative refractive index. IRS is often composed of a large amount of passive device elements, having low power consumption and low prices.
	
	Two main research directions are developed on IRS. One is employing IRS as transmitter or receiver to design a communication system. Another one is using IRS as wireless relay to reconfigure the communication environment. The mechanism that IRS can change the propogation direction of electromagnetic wave is the phase gradient distribution on IRS that is called The Generalized Snell's Law \cite{basic0}. In \cite{cui2}, the researchers presented a totally new wireless communication system using programmable metasurface as a transmitter. This new architecture did not need any complex algorithm of signal processing, any filter or power amplifier. They used a metasurface with $8\times32$ cells as transmitter to design this 8-phase shift-keying (8PSK) communication system working at 4.25 GHz, achieving an acceptable bit error rate and 6.144 Mbps transmission data rate. The authors in \cite{cui3} designed an architecture realizing quadrature phase shift keying (QPSK) over the air without channel coding with a $16\times8$ IRS. The phase of unit cells on this IRS was directly controlled by the baseband signal. This system achieved the transmission rate of 2.048 Mbps. They also found that through increasing the transmitting power by 5 dB, the performance of bit error rate can be the same with the conventional channel coding system. In \cite{Dai1}, the researchers designed a prototype with IRS having 256 elements and the phase shift resolution was 2-bit. This system had flexible software and modular hardware including a host for setting parameters, the universal software radio peripherals (USRPs) for baseband and radio frequency (RF) signal processing, as well as the IRS for signal transmission and reception.The performance evaluation showed that the antenna gain can be 21.7 dBi at 2.3 GHz and 19.1 dBi at 28.5 GHz.
	
	About the second main research direction, recent research mainly focused on the optimization design of the reflecting coefficients, channel estimation, application in physical security, etc. Reference \cite{basic1}, \cite{basic2} both proposed a simple two ray model to elaborate that the basic idea of the second research direction is eliminating the phase difference among the multipath signals, making the multipath fading mitigated. The difference between the distances that two rays travel is the main factor considered in \cite{basic1}. In \cite{basic2}, the authors did not only consider the distance factor but also the Doppler shift difference caused by the motion of transceivers. They drawed an important conclusion that through employing the real-time tunable IRS, the fading effect caused by the Doppler shift can be greatly mitigated. 
	
	The transmitting power and reflecting coefficients was corporately designed for a downlink multiple-input single-output (MISO) communication system with multiple users \cite{Huang1}. The maximized sum-rate was set as the object of this non-convex problem. They tackled this problem by using the alternating maximization and the majorization-minimization simultaneously and the numerical results showed that the promotion could be up to forty percent with the condition of no additional power. Reference \cite{Huang2} also considered the MISO scenario with multiple users but the optimization target was alternated as energy efficiency and the resolution of IRS is low, only 1-bit. The simulation result showed that with low resolution IRS, the system can also provide a higher energy efficiency than the conventional relay system. Authors in \cite{Huang3} studied a similar scenario as \cite{Huang2}. They designed the transmit power allocation and the phase shifts of IRS with two methods, gradient descent search and sequential fractional programming. The authors in \cite{SE} presented an idea of a joint optimization problem on spectrum efficiency. Reference \cite{WQ1} minimized the transmitting power at BS considering a MISO system with multiple users assisted by IRS bounded by users' signal to interference plus noise (SINR). Reference \cite{WQ2} considered the problem maximizing the achievable rate by designing the transmitting beamforming and IRS beamforming jointly under the condition that the phase and amplitude of the reflecting coefficients are related to each other. They derived this relationship through a circuit model. They gave out a suboptimal solution with a low-complexity based on the alternating optimization method. A communication scenario with multiple information decoding receivers (IDR) and energy harvesting receivers (EHR) was considered in \cite{WQ3}. 
	
	A physical safety problem employing IRS  was researched in \cite{WQ4}. Reference \cite{WQend} moved out the assumption of the continuous phase shifts and changed the problem into a mixed-integer nonlinear program. And a secrecy rate maximization problem under a severe scenario that the eavesdropping channel is stronger than the legitimate channel was solved through applying the alternating optimization and semidefinite relaxation methods. This article maximized the weighted sum of power at EHRs constrained by SINR at IDRs. These optimization problem are considered based on one assumption that the channel distribution information can be totally known at BS and IRS. In fact, most of these papers about the optimization problem on IRS assumed a simple channel distribution, such as Rayleigh distribution or Rician distribution. A more complicated distribution which is Nakagami's m-distribution was considered in \cite{CM1}. This paper also proposed a hardware model, a signal model and a path loss model. When it comes to the realization of IRS in the real world, narrowband channel model is not accurate any more. So some researchers were devoted to studying the channel estimation about IRS. Reference \cite{chesti1} proposed a channel estimation protocol based on the minimum mean squared error (MMSE) for a multi-user MISO system. The main idea of this protocol was separating the coherence time of the estimated channel into two parts, the former one is for channel estimation and the latter one is for transmission according to the estimation results. During the first period, all the elements are closed at the beginning and then the elements of IRS are opened one by one to estimate the corresponding channel information. Reference \cite{chesti2} proposed a channel estimation method according to the pilots received from the users and derived the channel estimation error in closed form. The influence on system performance caused by the phase error was studied in \cite{PERROR}. Reference \cite{chesti3} proposed a channel estimation method that is based on compressed sensing and deep learning. 
	
	Besides these two main directions, combining high frequency communication with IRS is also an attractive research field. For higher transmission rate, higher frequency band communication is an inexorable trend in the future \cite{5GCMsur}. With the communication frequency rising, the wavelength is getting shorter and the size of IRS can be smaller resulting in more convenient installation of IRS. Terahertz (THz) communication combined with IRS was studied in \cite{highband1}. Visible light communication with IRS was researched in \cite{highband2}.
	
	An appropriate channel model is very important for system design and performance evaluation. There were also few research achievements coming out about channel modeling with IRS recently. Reference \cite{basic0} derived the path loss expression for the far-field case from the point of view of physical propogation. Reference \cite{PL1} demonstrated a generalized path loss expression which is correct but complicated for both far-field and near-field cases. Then the authors started from this to the other three special cases which are far-field case, near-field case, and broadcasting case to derive the expressions. All the situations did not consider the direct link between the transmitter and the receiver. They all assumed that there exists a line of sight (LoS) component. Moreover this article also derived the beamforming design on IRS making the path loss minimum. In \cite{CM1}, the authors combined the 5G channel model with IRS to demonstrate that IRS and decode-and-forward (DF) relays can complement each other's shortages and IRS can not replace DF relays completely. In \cite{CM2}, a channel model based on the standardized channel model was proposed considering two different scenarios which are indoor and outdoor. This article set the path loss for every ray respectively. It did not support the MIMO application and movements. Moreover the scatterers' distribution was also simple.
	
	To the best of our knowledge, an appropriate channel model for IRS is still missing in this area. This paper proposes a geometry based stochastic channel model (GBSM), which supports the movements of transceivers and clusters, and the evolution of clusters in space domain. The reflecting coefficients design based on the minimum path loss is also considered. The steering vector is set at BS to cooperate with IRS. Finally, the simulation results and analytical results are compared to prove the accuracy of this channel model.
	
	The remainder of this paper is organized as follows. Section~\uppercase\expandafter{\romannumeral2} describes the system model, main assumptions, and some important technical terms in channel modeling. Section~\uppercase\expandafter{\romannumeral3} describes the channel model in details and we derive the channel impulse response (CIR). In Section~\uppercase\expandafter{\romannumeral4}, the statistical properties of the proposed channel model are calculated. In Section~\uppercase\expandafter{\romannumeral5}, the simulation results are presented. Finally, we draw conclusions in Section~\uppercase\expandafter{\romannumeral6}.

\section{System Model}
\begin{figure}[tb]
	\centerline{\includegraphics[width=0.5\textwidth]{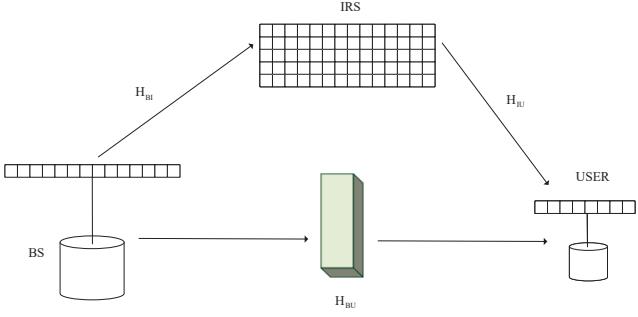}}
	\caption{A wireless communication scenario using IRS.}
	\label{basic scenario}
\end{figure}	
    Let us consider a MIMO communication scenario employing IRS as illustrated in Fig. \ref{basic scenario}. BS is equipped with $M_\text{B}$ antennas. USER is equipped with $M_\text{U}$ antennas and IRS employs $M_{xy}$ passive reflecting elements. Here $M_{xy}$ is calculated as $M_x\times M_y$, where $M_x$ and $M_y$ are the numbers of elements on horizontal and vertical directions of IRS, respectively. BS and USER are both uniform linear arrays. IRS is a uniform plane array. All the antenna elements are omnidirectional. There are three sub-channels in this communication system i.e., the channel between BS and IRS, the channel between IRS and USER, and the channel between BS and USER. The whole channel coefficients matrix is denoted as $\textbf{H}_\text{total}$. Then, the channel coefficients matrix can be expressed as \cite{WCSP}
\begin{equation}\label{total channel}
\begin{split}
    \textbf{H}_\text{total}&=(\textbf{H}_{\text{IU}}\mathbf{\Phi}\textbf{H}_{\text{BI}}+\textbf{H}_{\text{BU}})\textbf{f}\\
    &=(\sqrt{SF_{\text{BI}}SF_{\text{IU}}PL_{\text{BIU}}}\textbf{h}_{\text{IU}}\mathbf{\Phi}\textbf{h}_{\text{BI}}\\
    &+\sqrt{SF_{\text{BU}}PL_{\text{BU}}}\textbf{h}_{\text{BU}})\textbf{f}
\end{split}
\end{equation}
	where $\textbf{H}_{\text{BI}}\in \mathbb{C}^{M_{xy}\times M_{B}}$, $\textbf{H}_{\text{IU}}\in \mathbb{C}^{M_{U}\times M_{xy}}$, and $\textbf{H}_{\text{BU}}\in \mathbb{C}^{M_{U}\times M_{B}}$ are the corresponding channel coefficients matrices, respectively. $\textbf{H}$ and $\textbf{h}$ are the channel coefficients matrices consisting of large scale fading and not consisting of large scale fading respectively. The reflecting coefficients matrix of IRS is presented as $\mathbf{\Phi}$, which is a diagonal matrix whose dimension is $M_{xy}\times M_{xy}$. In this article, we mainly consider two kinds of large scale fading, shadowing effect and path loss. $SF_{\text{BI}}$, $SF_{\text{BU}}$, and $SF_{\text{IU}}$ are corresponding log normal random variables of different sub-channels to model the shadowing fading effect. $PL_{\text{BU}}$ denotes the path loss of the sub-channel which is between BS and user. $PL_{\text{BIU}}$ is the path loss of the cascaded channel assisted by IRS. $\textbf{f}$ is the steering vector of BS. The calculation of all the above denotations will be discussed later.

	We will also describe some important technical terms in channel modeling which will be discussed in the third part of subsection A of Section~\uppercase\expandafter{\romannumeral3}. As we all know, the transmitted electromagnetic signal will experience multiple physical interactions with different objects in the propagation environment and finally be received and summed by the receiver. This effect is known as multi-path effect. The interacting objects are named as scatterers. One scatterer corresponds to one ray. The rays reflected or scattered by these objects are called multi-path components (MPCs). A cluster is a group of MPCs with similar properties such as propogation delay, ray power, and angle of arrival or departure. Different propogation ways between two antenna elements of two communication terminals are called paths. In the proposed twin cluster channel model, one path corresponds to a pair of clusters. The propagation between two clusters or between one cluster and one communication terminal is called one bounce.

\section{A Novel 3D IRS MIMO GBSM}
\subsection{Description of the Channel Model}
\begin{figure}[tb]
	\centerline{\includegraphics[width=0.5\textwidth]{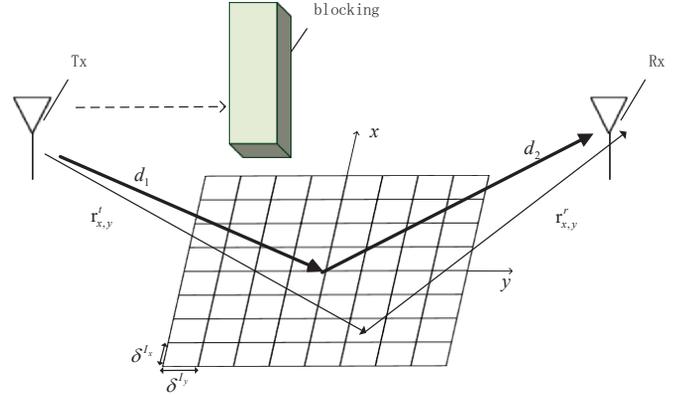}}
	\caption{A scenario for deriving reflecting coefficients.}
	\label{single antenna scenario}
\end{figure}

\subsubsection{The Reflecting Coefficient Matrix Setting}
	Indeed, the design of reflecting coefficients matrix is a complex problem. In specialized optimization problem, it is usually solved through some complex algorithms. To simplify the analysis of the channel model, we consider a special issue for the system with IRS. To illustrate the idea, we need to consider a situation of the transmitter (Tx) and the receiver (Rx) employed with single antenna shown in Fig.\ref{single antenna scenario} first. Another important assumption is that the directions of peak radiation of the transmitting and receiving antennas point to the center of IRS. Then we will use the steering vector to expand the conclusion to the situation that Tx and Rx are equipped with multiple antennas to support the proposed channel model that Tx and Rx are equipped with multiple omnidirectional antennas.
	
\begin{figure}[tb]
	\centerline{\includegraphics[width=0.5\textwidth]{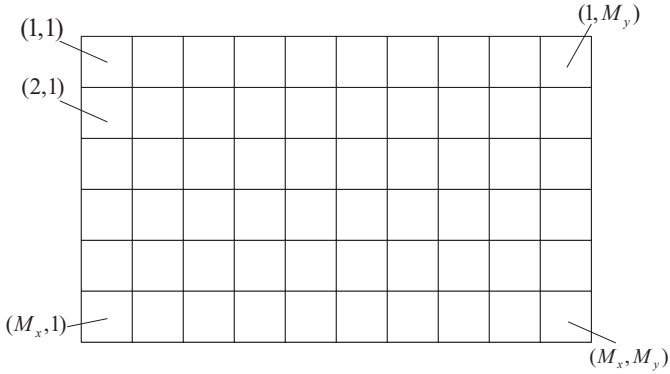}}
	\caption{The ordered pair for locating elements on 2D IRS.}
	\label{locationindex}
\end{figure}

	In Fig.\ref{single antenna scenario}, Tx and Rx are both employed with one antenna. Here we need to introduce the two approaches of locating the elements on IRS. Because this will be used to bridge the gap between the index of the reflecting coefficients matrix and the index of IRS. $(x,y)$ is an ordered pair meaning the location index of each element on IRS. $M_x$ and $M_y$ are the numbers of elements on horizontal and vertical directions of IRS. For a uniform linear array that is one dimensional (1D). We only need one number to position each element. To reduce the dimension of the matrix, we can transform the two dimensional (2D) index pair $(x,y)$ into this one number $r$ through traversing IRS by row. The transformation expression is 
\begin{equation}
	r=(x-1)M_{y}+y.
\end{equation}
	The transformation from $r$ to $(x,y)$ is shown as
\begin{equation}\label{xy_r1}
	x=r // M_{y}+1
\end{equation}
and
\begin{equation}\label{xy_r2}
	y=\text{mod}(r,M_{y})  
\end{equation}
	where $//$ means the integer division and $\text{mod}(\cdot)$ denotes the delivery operator. The approach we locate each element is shown in Fig. \ref{locationindex}. In the discussion of channel coefficients matrices below, we will mainly use one index to locate the elements on IRS.
	
	The distances between Tx with single antenna and each element on IRS are denoted as $r_{x,y}^t$. The distances between $\text{R}_\text{X}$ and each element are presented as $r_{x,y}^r$. $\delta^{\text{I}_{x}}$ and $\delta^{\text{I}_{y}}$ are the interval of the elements of IRS on two directions. $\phi_{x,y}$ is the phase of the corresponding element on IRS. With these premises, we can get the relationship between the received power $P_r$ and the transmitted power $P_t$ shown as\cite{PL1}
\begin{equation}\label{ptpr}
	P_r=P_t\frac{\delta^{\text{I}_{x}}\delta^{\text{I}_{y}}\lambda^2}{64{\pi}^3}\left|\sum_{x=1}^{M_x}\sum_{y=1}^{M_y}\frac{e^{\frac{-j(2\pi(r_{x,y}^r+r_{x,y}^t)-\lambda\phi_{x,y})}{\lambda}}}{r_{x,y}^{r}r_{x,y}^t}\right|^2
\end{equation}
	where $\lambda$ is the wavelength.
	From (\ref{ptpr}), we can easily see that the received power is maximized when the phase shifts fulfill 
\begin{equation}\label{BIU_condition}
	\phi_{x,y}=\text{mod}(\frac{2\pi(r_{x,y}^r+r_{x,y}^t)}{\lambda},2\pi).
\end{equation}
	More detailed derivation can be seen in \cite{PL1}. So the reflecting coefficients matrix can be obtained according to (\ref{BIU_condition}). Here we need to use one index $r$ to replace $(x,y)$to locate one element. The matrix is thus shown as
\begin{equation}
	\mathbf{\Phi}=\text{diag}(e^{j\phi_1},e^{j\phi_2}...e^{j\phi_r}...e^{j\phi_{M_{xy}}})
\end{equation}
	where $\text{diag}(\cdot)$ means the operation of generating a diagonal matrix through arranging the numbers in the embrace as its elements on the leading diagonal line in order. When it comes to implement IRS in the real world, we must consider the situation that the phase shifts are with finite resolution. So the values of elements in the matrix $\mathbf{\Phi}$ will be selected in a set with a finite number of values. Here we consider that IRS has a 2-bit quantization. The phase shifts set is $\left\{\frac{\pi}{4},\frac{3\pi}{4},\frac{5\pi}{4},\frac{7\pi}{4}\right\}$. We just need to replace the value of $\left\{\phi_1,\phi_2...\phi_r...\phi_{M_{xy}}\right\}$ with the value in this set nearest by them.
	
\begin{figure}[tb]
	\centerline{\includegraphics[width=0.5\textwidth]{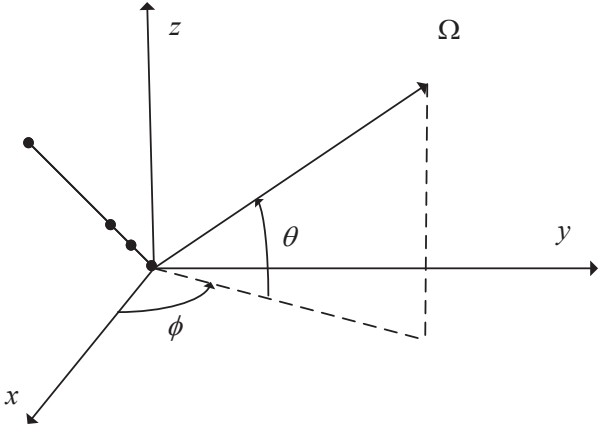}}
	\caption{Steering vector setting.}
	\label{steeringvector}
\end{figure}

	The channel model we propose needs to support the situation of Tx and Rx employed with multiple antennas. To follow the former conclusion suitable for single antenna situation, we should guarantee the peak radiation of all the antennas still points to the center of IRS. According to \cite{steering1}, we can add extra phase shifts for every element on Tx to achieve this goal. These phase shifts on Tx generate the steering vector. The steering vector can be expressed as \cite{steering1}
\begin{equation}\label{steering}
	\mathbf{f}=\left[c_{1}(\mathbf{\Omega}),c_{2}(\mathbf{\Omega})...c_{m}(\mathbf{\Omega})...c_{M_T}(\mathbf{\Omega})\right]
\end{equation}
	where $\mathbf{\Omega}$ means the direction pointing to the center of IRS, and $c_{m}(\mathbf{\Omega})$ is the extra phase shift coefficients of the $m\text{th}$ antenna. And the coefficients are calculated as \cite{steering1}
\begin{equation}\label{steering2}
	c_{m}(\mathbf{\Omega})=exp(j2\pi \lambda^{-1} \left\langle\textbf{e}(\mathbf{\Omega}),\textbf{r}_m\right\rangle  +j2\pi\nu_{l}t)
\end{equation}
	where $\nu_{l}$ is the Doppler shift. $\left\langle \cdot \right\rangle$ denotes the dot product of two vectors. $\textbf{e}(\mathbf{\Omega})$ and $\textbf{r}_m$ are the unit vector of the departure direction and the vector related to the antenna interval. $(x_{m},y_{m},z_{m})$ is the Cartesian value of the $m\text{th}$ element on BS. The spatial relationship is shown in Fig. \ref{steeringvector}. $d_{x}$, $d_{y}$ and $d_{z}$ are usually valued as half of the wavelength. The unit vectors of the departure direction is calculated as
\begin{equation}
	\textbf{e}(\mathbf{\Omega})=(\text{cos}\theta \text{cos}\Phi, \text{cos}\theta \text{sin}\Phi, \text{sin}\theta).
\end{equation}
	The vector $\textbf{r}_{m}$ is calculated as \cite{steering2}
\begin{equation}
	\textbf{r}_{m}=(x_m-d_x,y_m-d_y,z_m-d_z).
\end{equation}

\subsubsection{Large Scale Fading Channel Coefficient}
	About large scale fading, we consider two parts of it. The first one is shadowing fading, the other one is path loss. To make the channel model more general, we model these two properties for every sub-channel. Under some situations, different sub-channels may share the same large scale fading. Shadowing effect is mainly caused by the obstruction between Tx and Rx. In (\ref{total channel}), we use $SF_{\text{BI}}$, $SF_{\text{BU}}$, and $SF_{\text{IU}}$ to present the corresponding shadowing fading effect of different sub-channels. The probability density function can be written as
\begin{equation}
	p_{SF}(x)=\frac{2}{x\sigma_{SF}\text{ln}10/10}\text{exp}(-\frac{(10\text{log}_{10}x^{2}-\mu)^{2}}{2{\sigma_{SF}}^{2}})
\end{equation}
	where $\sigma_{SF}$ is shadowing standard derivation determined by the scenario, $\mu$ is the mean value of this random variable with the unit of $\text{dB}$.

	About path loss, we classify it into two kinds, one is path loss of the sub-channel between BS and USER, the other one is path loss of the cascaded channel assisted by IRS. $PL_{\text{BU}}$ is denoted as path loss of the sub-channel which is between BS and user. $PL_{\text{BIU}}$ is the path loss of the cascaded channel assisted by IRS. For the former one, we employ the path loss expression of the standardized QuaDRiGa channel model \cite{Quadriga} as
\begin{equation}
	PL_{\text{BU}}^{[\text{dB}]}=-A\cdot \text{log}_{10}d_{[\text{km}]}-B-C\cdot \text{log}_{10}f_{[\text{GHz}]}
\end{equation}
	where $A, B,$ and $C$ are the parameters determined by the communication scenario. Through substituting the optimized reflecting coefficients into (\ref{ptpr}), we can obtain the path loss of the sub-channel assisted by IRS as 

	\begin{equation}
	PL_{\text{BIU}}=\frac{\delta^{I_{x}}\delta^{I_{x}}\lambda^{2}\left| \sum_{x=1}^{M_{x}}\sum_{y=1}^{M_{y}}\frac{1}{r_{x,y}^{t}r_{x,y}^{r}}\right|^2 }{64\pi^{3}}.
	\end{equation}

\begin{figure}[tb]
	\centerline{\includegraphics[width=0.5\textwidth]{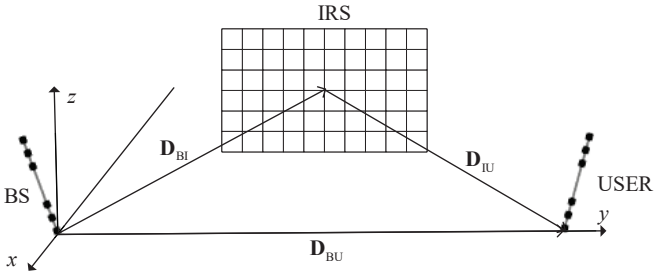}}
	\caption{The vector relationship among three terminals.}
	\label{pointingvector}
\end{figure}
\subsubsection{Small Scale Fading Channel Coefficient Matrix}
	To make the channel model more accurate and more general, we use a GBSM to model it. This kind of channel model is composed of deterministic parts and stochastic parts. The former one mainly includes some system parameters such as the location information of the transceivers, the velocities of the mobile user or the BS, the scenario and so on. The latter one usually includes the scatterers' and the clusters' distribution. 
	
    We will firstly introduce how to determine the basic location relationship among three communication terminals which are BS, IRS, and USER at initial time. Here we use three vectors pointing from one of them to another. They are denoted as $\textbf{D}_{\text{BI}}$, $\textbf{D}_{\text{IU}}$, and $\textbf{D}_{\text{BU}}$. In more detail, $\textbf{D}_{\text{BI}}$ is the vector pointing from the first element to the center of IRS at initial time. $\textbf{D}_{\text{IU}}$ is the vector pointing from the center of IRS to the first element of USER at initial time. $\textbf{D}_{\text{BU}}$ is the vector pointing from the first element of BS to the first element of USER at initial time. The pointing relationship is shown in Fig. \ref{pointingvector}. Actually we only need to set any two of them to determine the whole location relationship because three vectors are constrained by the expression shown as
\begin{equation}
	\textbf{D}_{\text{BU}}=\textbf{D}_{\text{BI}}+\textbf{D}_{\text{IU}}.
\end{equation}
\begin{figure}[tb]
	\centerline{\includegraphics[width=0.5\textwidth]{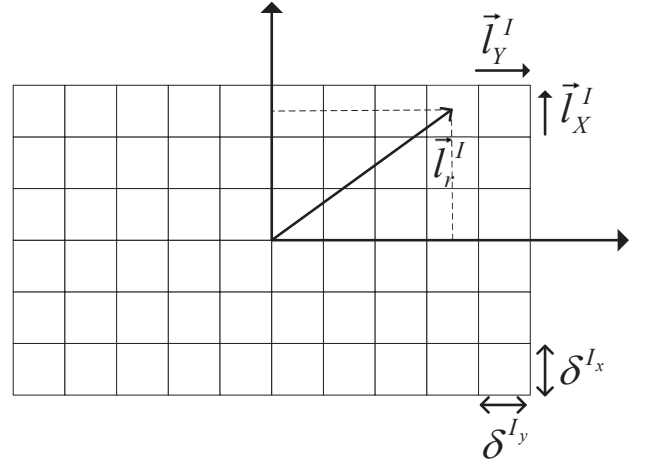}}
	\caption{The method of finding the elements through the center.}
	\label{IRSvector}
\end{figure}
	The method of placing three terminals is determined by some angle parameters and interval parameters. Let $\beta_A^\text{B}$ and $\beta_E^\text{B}$ be the azimuth angle and elevation angle of BS linear array. $\beta_A^\text{U}$ and $\beta_E^\text{U}$ are the azimuth angle and the elevation angle of USER linear array, respectively. $\beta_A^{\text{I}_x}$ and $\beta_A^{\text{I}_y}$ are azimuth angles of two extending directions of IRS. Elevation angles of two extending directions of IRS are denoted as $\beta_E^{\text{I}_x}$ and $\beta_E^{\text{I}_y}$. $\delta^{\text{B}}$, $\delta^{\text{U}}$, $\delta^{\text{I}_x}$ and $\delta^{\text{I}_y}$ represent the antenna elements interval of BS, USER, and two extending directions of IRS respectively. The $q\text{th}$, $r\text{th}$, and $p\text{th}$ elements on BS, IRS, and USER are denoted as $A_q^\text{B}$, $A_r^\text{I}$, and $A_p^\text{U}$. Here we use one index approach to locate the elements on IRS which is equivalent to the ordered pair index. According to these parameters, we can determine the vectors pointing from the first element or the center to the other elements. They are respectively denoted as $\textbf{l}_{q}^{\text{B}}$, $\textbf{l}_{r}^{\text{I}}$, and $\textbf{l}_{p}^{\text{U}}$. $\textbf{l}_{q}^{\text{B}}$ is the vector from $A_{1}^\text{B}$ to $A_q^\text{B}$. $\textbf{l}_{r}^{\text{I}}$ is the vector from the center of IRS to $A_r^\text{I}$. $\textbf{l}_{p}^{\text{U}}$ is the vector from $A_{1}^\text{U}$ to $A_p^\text{U}$. The vector $\textbf{l}_{r}^{\text{I}}$ is shown in Fig. \ref{IRSvector}. They are calculated as
\begin{equation}
	\textbf{l}_{q}^{\text{B}}=(q-1)\delta^{\text{B}}[\text{cos}\beta_{E}^{\text{B}}\text{cos}\beta_{A}^{\text{B}}, \text{cos}\beta_{E}^{\text{B}}\text{sin}\beta_{A}^{\text{B}}, \text{sin}\beta_{E}^{\text{B}}]
\end{equation}
\begin{equation}
	\textbf{l}_{p}^{\text{U}}=(p-1)\delta^{\text{U}}[\text{cos}\beta_{E}^{\text{U}}\text{cos}\beta_{A}^{\text{U}}, \text{cos}\beta_{E}^{\text{U}}\text{sin}\beta_{A}^{\text{U}}, \text{sin}\beta_{E}^{\text{U}}]
\end{equation}
and
\begin{equation}
	\textbf{l}_{r}^{\text{I}}=(\frac{M_{x}+1}{2}-x)\textbf{l}_{X}^{\text{I}}+(y-\frac{M_{y}+1}{2})\textbf{l}_{Y}^{\text{I}}
\end{equation}
	where $x$ and $y$ can be transformed from $r$ according to (\ref{xy_r1}) and (\ref{xy_r2}). $\textbf{l}_{X}^{\text{I}}$ and $\textbf{l}_{Y}^{\text{I}}$ are calculated as
\begin{equation}
	\textbf{l}_{X}^{\text{I}}=\delta^{I_x}[\text{cos}\beta_{E}^{I_{x}}\text{cos}\beta_{A}^{I_{x}}, \text{cos}\beta_{E}^{I_{x}}\text{sin}\beta_{A}^{I_{x}}, \text{sin}\beta_{E}^{I_{x}}]
\end{equation}
and 
\begin{equation}
	\textbf{l}_{Y}^{\text{I}}=\delta^{I_y}[\text{cos}\beta_{E}^{I_{y}}\text{cos}\beta_{A}^{I_{y}}, \text{cos}\beta_{E}^{I_{y}}\text{sin}\beta_{A}^{I_{y}}, \text{sin}\beta_{E}^{I_{y}}].
\end{equation}

\begin{figure}[tb]
	\centerline{\includegraphics[width=0.5\textwidth]{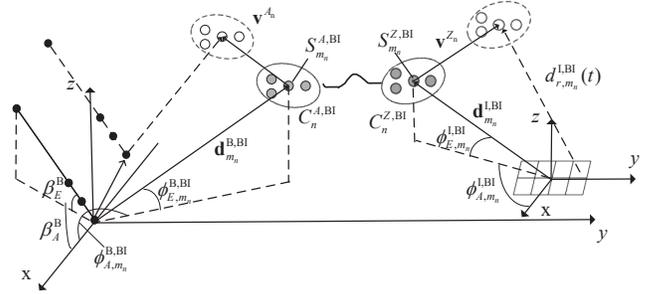}}
	\caption{The GBSM for IRS channel.}
	\label{twincluster}
\end{figure}

	The total number of clusters between $A_q^{\text{B}}$ and $A_r^{\text{I}}$ at time instant $t$ is denoted as $N_{qr}^{\text{BI}}(t)$. Similarly the number of clusters between $A_q^{\text{B}}$ and $A_p^{\text{U}}$ at time instant $t$ is denoted as $N_{qp}^{\text{BU}}(t)$ and the number of clusters between $A_r^{\text{I}}$ and $A_p^{\text{U}}$ at time instant $t$ is denoted as $N_{rp}^{\text{IU}}(t)$. After interacting with the scatterers, the transmitted signal will be changed in power or in phase. These changed signal rays will be received and summed by the receiver together. Here we only consider the two clusters which are nearest by Tx and Rx. In the BS-IRS sub-channel, the $n\text{th}$ cluster near BS is denoted as $C_{n}^{A,\text{BI}}$, the one near IRS is denoted as $C_{n}^{Z,\text{BI}}$. Similarly, in the other two sub-channels, the clusters near the transmitters and near the receivers are denoted as $C_{n}^{A,\text{IU}}$, $C_{n}^{Z,\text{IU}}$, $C_{n}^{A,\text{BU}}$, and $C_{n}^{Z,\text{BU}}$. The other bounces between this two clusters are modeled as one virtual link which will be discussed later. There are some scatterers in one cluster. Correspondingly, in BS-IRS sub-channel, the $m\text{th}$ scatterers in $C_{n}^{A,\text{BI}}$ and $C_{n}^{Z,\text{BI}}$ are denoted as $S_{m_n}^{A,\text{BI}}$ and $S_{m_n}^{Z,\text{BI}}$. For the paths of the other two sub-channels, the scatterers are denoted as $S_{m_n}^{A,\text{IU}}$, $S_{m_n}^{Z,\text{IU}}$, $S_{m_n}^{A,\text{BU}}$, and $S_{m_n}^{Z,\text{BU}}$.
		\begin{table*}[t]
		\centering
		\setlength{\belowcaptionskip}{0.3cm}
		\caption{Definitions of main parameters for the proposed IRS-assisted channel model.}
		\begin{tabular}{|l|l|}			
			\hline
			\textbf{Parameters}&\textbf{Definitions}\\
			\hline
			$\text{D}_{\text{BI}}$,$\text{D}_{\text{IU}}$,$\text{D}_{\text{BU}}$&The pointing vectors between the first element of BS/USER and the center of IRS
			\\\hline
			$M_{\text{B}}$,$M_{\text{U}}$,$M_{x}$,$M_{y}$&The numbers of elements on BS, IRS, and two extending directions of IRS
			\\\hline
			$A_{q}^{\text{B}}$, $A_{r}^{\text{I}}$, $A_{p}^{\text{U}}$&The $q\text{th}$/$r\text{th}$/$p\text{th}$ element on BS/IRS/USER
			\\
			\hline
			$\textbf{l}_{q}^{\text{B}}$, $\textbf{l}_{p}^{\text{U}}$, $\textbf{l}_{r}^{\text{I}}$& The vectors pointing from the first element or the center to the other elements\\
			\hline
			$\delta^{\text{B}},\delta^{\text{U}},\delta^{\text{I}_x},\delta^{\text{I}_y}$&Intervals between the elements on different communication terminals
			\\\hline
			$\beta_{A}^{\text{B}/\text{I}_x/\text{I}_y/\text{U}}$,$\beta_{E}^{\text{B}/\text{I}_x/\text{I}_y/\text{U}}$&The elevation angles and the azimuth angles of BS/IRS/USER layout
			\\\hline
			$C_{n}^{A,\text{BI/IU/BU}}$,$C_{n}^{Z,\text{BI/IU/BU}}$&The first bounce and the last bounce clusters of the corresponding sub-channel
			\\\hline
			$v^{\text{B}}(t)$,$v^{\text{U}}(t)$,$v^{A_n}(t)$,$v^{Z_n}(t)$&The absolute value of the velocity of BS, USER, and clusters
			\\\hline
			$\alpha^{\text{B}}(t)$,$\alpha^{\text{U}}(t)$,$\alpha^{A_n}(t)$,$\alpha^{Z_n}(t)$&The azimuth angles of 2D velocity vectors
			\\\hline
			$\phi_{A,m_n}^{\text{B,BI/B,BU/I,IU}}$,$\phi_{E,m_n}^{\text{B,BI/B,BU/I,IU}}$&The AAoD and EAoD of the $m\text{th}$ ray in the $n\text{th}$ cluster at initial time.
			\\\hline
			$\phi_{A,m_n}^{\text{I,BI/U,BU/U,IU}}$,$\phi_{E,m_n}^{\text{I,BI/U,BU/U,IU}}$&The AAoA and EAoA of the $m\text{th}$ ray in the $n\text{th}$ cluster at initial time.
			\\\hline
			$d_{m_n}^{\text{B,BI/B,BU/I,IU}}$, $d_{m_n}^{\text{I,BI/U,BU/U,IU}}$&The distance between $S_{m_n}$ and the first element at initial time.
			\\\hline
			$d_{q/r,m_n}^{\text{B,BI/B,BU/I,IU}}(t)$, $d_{r/p,m_n}^{\text{I,BI/U,BU/U,IU}}(t)$&The distance between $S_{m_n}$ and any element at time instant $t$.
			\\\hline
		\end{tabular}
		\label{tab1}
	\end{table*}	
	We could determine relative position relationship between scatterers and any antenna elements through using one distance parameter and two angle parameters. The three parameters can determine the relative position relationship between scatterers and the first element of the linear array or the center of IRS. Let $\phi_{A,m_n}^{\text{B,BI}}$ and $\phi_{E,m_n}^{\text{B,BI}}$ be azimuth angle of departure (AAoD) and elevation angle of departure (EAoD) of the $m\text{th}$ ray at initial time. Similarly, $\phi_{A,m_n}^{\text{I,BI}}$ and $\phi_{E,m_n}^{\text{I,BI}}$ are the azimuth angle of arrival (AAoA) and the elevation angle of arrival (EAoA) of the ray coming from $S_{m_n}^{Z,\text{BI}}$ received by the center of IRS. The angle parameters for the other two sub-channels are denoted as $\phi_{A,m_n}^{\text{I,IU}}$, $\phi_{E,m_n}^{\text{I,IU}}$, $\phi_{A,m_n}^{\text{U,IU}}$, $\phi_{E,m_n}^{\text{U,IU}}$, $\phi_{A,m_n}^{\text{B,BU}}$, $\phi_{E,m_n}^{\text{B,BU}}$, $\phi_{A,m_n}^{\text{U,BU}}$, and $\phi_{E,m_n}^{\text{U,BU}}$. The distances between any scatterer and any antenna element at initial time instant are denoted as $d_{q,m_n}^{\text{B,BI}}$, $d_{r,m_n}^{\text{I,BI}}$, $d_{q,m_n}^{\text{B,BU}}$, $d_{p,m_n}^{\text{U,BU}}$, $d_{r,m_n}^{\text{I,IU}}$, and $d_{p,m_n}^{\text{U,IU}}$. With the motion of terminals and clusters, the distances will change and influence the calculation of the delay. For example, $d_{q,m_n}^{\text{B,BI}}$ will change into $d_{q,m_n}^{\text{B,BI}}(t)$ at time instance $t$. The main parameters and the twin cluster channel model are illustrated in Fig. \ref{twincluster}.
	
	In the proposed model, the mobilities of BS, clusters, and USER are supported. The speed vectors of BS, USER and clusters at different time instances are denoted as $\mathbf{v}^{\text{B}}(t)$, $\mathbf{v}^{\text{U}}(t)$, $\mathbf{v}^{A_n}(t)$ and $\mathbf{v}^{Z_n}(t)$. We assume that they can only move in a 2D plane. So only one angle can determine the directions of their movements which are presented as $\alpha^{\text{B}}(t)$, $\alpha^{\text{U}}(t)$, $\alpha^{A_n}(t)$, $\alpha^{Z_n}(t)$. The main parameters are summarized in Table~\ref{tab1}.

	Then we will illustrate how to generate the small scale fading channel coefficients matrix. Without loss of generality, we mainly use the BS-IRS sub-channel to demonstrate this procedure. The elements in the small scale fading matrix $\textbf{h}_{\text{BI}}$ are denoted as $h_{qr,\text{BI}}(t,\tau)$ that represents the channel coefficients between $A_q^\text{B}$ and $A_r^\text{I}$ at time instant $t$. It is calculated as
\begin{equation}\label{channelresponse_BI}
	h_{qr,\text{BI}}(t,\tau)=\sqrt{\frac{K}{K+1}}h_{qr,\text{BI}}^{L}(t,\tau)+\sqrt{\frac{1}{K+1}}h_{qr,\text{BI}}^{N}(t,\tau).
\end{equation}
	Similarly, we can obtain the CIR between $A_r^\text{I}$ and $A_p^\text{U}$ and the CIR between $A_q^\text{B}$ and $A_p^\text{U}$. They are shown as 
\begin{equation}\label{channelresponse_IU}
	h_{rp,\text{IU}}(t,\tau)=\sqrt{\frac{K}{K+1}}h_{rp,\text{IU}}^{L}(t,\tau)+\sqrt{\frac{1}{K+1}}h_{rp,\text{IU}}^{N}(t,\tau)
\end{equation}
and
\begin{equation}\label{channelresponse_BU}
	h_{qp,\text{BU}}(t,\tau)=\sqrt{\frac{K}{K+1}}h_{qp,\text{BU}}^{L}(t,\tau)+\sqrt{\frac{1}{K+1}}h_{qp,\text{BU}}^{N}(t,\tau).
\end{equation}
	Obviously, it is obtained by summing two weighed components. The former one is the line of sight (LoS) component, and the latter one is non-line of sight (NLoS) component. Their weights are determined by the Rician factor $K$. According to the previous assumption of omnidirectional antenna, the antenna pattern is ignored in the following calculation. We calculate the NLoS component as\cite{Bianji}

\begin{equation}\label{NCIR_BI}
\begin{split}
	h_{qr,\text{BI}}^{N}(t,\tau)=&\sum_{n=1}^{N_{qr,\text{BI}}(t)}\sum_{m_{n}=1}^{M_n}\sqrt{P_{qr,\text{BI},m_n}(t)}\\
	&e^{j2\pi f_{c}\tau_{qr,,\text{BI},m_n}(t)}\cdot\delta(\tau-\tau_{qr,\text{BI},m_n}(t))
\end{split}
\end{equation}

	\begin{equation}\label{NCIR_IU}
		\begin{split}
			h_{rp,\text{IU}}^{N}(t,\tau)=&\sum_{n=1}^{N_{rp,\text{IU}}(t)}\sum_{m_{n}=1}^{M_n}\sqrt{P_{rp,\text{IU},m_n}(t)}\\
			&e^{j2\pi f_{c}\tau_{rp,\text{IU},m_n}(t)}\cdot\delta(\tau-\tau_{rp,\text{IU},m_n}(t))
		\end{split}
	\end{equation}

and

	\begin{equation}\label{NCIR_BU}
		\begin{split}
			h_{qp,\text{BU}}^{N}(t,\tau)=&\sum_{n=1}^{N_{qp,\text{BU}}(t)}\sum_{m_{n}=1}^{M_n}\sqrt{P_{qp,\text{BU},m_n}(t)}\\
			&e^{j2\pi f_{c}\tau_{qp,\text{BU},m_n}(t)}\cdot\delta(\tau-\tau_{qp,\text{BU},m_n}(t))
		\end{split}
	\end{equation}

	where $\tau_{qr,\text{BI},m_n}(t)$ denotes the delay of the propagation link, $A_q^{\text{B}}\to S_{m_n}^{A,\text{BI}}\to \text{virtual link}\to S_{m_n}^{Z,\text{BI}}\to A_r^{\text{I}}$. Similarly, $\tau_{rp,\text{IU},m_n}(t)$ and $\tau_{qp,\text{BU},m_n}(t)$ denote the corresponding propogation links, $A_r^{\text{I}}\to S_{m_n}^{A,\text{IU}}\to \text{virtual link}\to S_{m_n}^{Z,\text{IU}}\to A_p^{\text{U}}$, and $A_q^{\text{B}}\to S_{m_n}^{A,\text{BU}}\to \text{virtual link}\to S_{m_n}^{Z,\text{BU}}\to A_p^{\text{U}}$. $P_{qr,\text{BI},m_n}(t)$ is the power of the ray interacting with $S_{m_n}^{A,\text{BI}}$ and $S_{m_n}^{Z,\text{BI}}$. $P_{rp,\text{IU},m_n}(t)$ is the power of the ray interacting with $S_{m_n}^{A,\text{IU}}$ and $S_{m_n}^{Z,\text{IU}}$. $P_{qp,\text{BU},m_n}(t)$ is the power of the ray interacting with $S_{m_n}^{A,\text{BU}}$ and $S_{m_n}^{Z,\text{BU}}$. $\delta(\cdot)$ denotes Dirac delta function. The ray delays are calculated as\cite{Bianji}
\begin{equation}\label{nlosdelay_BI}
	\tau_{qr,\text{BI},m_n}(t)=d_{qr,\text{BI},m_n}(t)/c+\tau_{n,\text{BI}}^{v}
\end{equation}
\begin{equation}\label{nlosdelay_IU}
	\tau_{rp,\text{IU},m_n}(t)=d_{rp,\text{IU},m_n}(t)/c+\tau_{n,\text{IU}}^{v}
\end{equation}
and
\begin{equation}\label{nlosdelay_BU}
	\tau_{qp,\text{BU},m_n}(t)=d_{qp,\text{BU},m_n}(t)/c+\tau_{n,\text{BU}}^{v}
\end{equation}
	where $c$ is the light speed. $\tau_{n,\text{BI}}^{v}$, $\tau_{n,\text{IU}}^{v}$, and $\tau_{n,\text{BU}}^{v}$ represent the virtual link delay between $C_{n}^{A,\text{BI}}$ and $C_{n}^{Z,\text{BI}}$, the virtual link delay between $C_{n}^{A,\text{IU}}$ and $C_{n}^{Z,\text{IU}}$, and the virtual link delay between $C_{n}^{A,\text{BU}}$ and $C_{n}^{Z,\text{BU}}$, which are all modeled as a random variable that follows an exponential distribution determined by the scenario. $d_{qr,\text{BI},m_n}(t)$ is the sum of two propagation distances, $A_q^{\text{B}}\to S_{m_n}^{A,\text{BI}}$, and $S_{m_n}^{Z,\text{BI}}\to A_r^{\text{I}}$. $d_{rp,\text{IU},m_n}(t)$ is the sum of two propagation distances, $A_r^{\text{I}}\to S_{m_n}^{A,\text{IU}}$, and $S_{m_n}^{Z,\text{IU}}\to A_p^{\text{IU}}$. $d_{qp,\text{BU},m_n}(t)$ is the sum of two propagation distances, $A_q^{\text{B}}\to S_{m_n}^{A,\text{BU}}$, and $S_{m_n}^{Z,\text{BU}}\to A_p^{\text{U}}$. They are calculated as 
\begin{equation}
	d_{qr,\text{BI},m_n}(t)=\left|\left|\textbf{d}_{q,m_n}^{\text{B,BI}}(t)\right| \right|+\left|\left| \textbf{d}_{r,m_n}^{\text{I,BI}}(t) \right| \right|
\end{equation}
\begin{equation}
	d_{rp,\text{IU},m_n}(t)=\left|\left|\textbf{d}_{r,m_n}^{\text{I,IU}}(t)\right| \right|+\left|\left| \textbf{d}_{p,m_n}^{\text{U,IU}}(t) \right| \right|
\end{equation}
and
\begin{equation}
	d_{qp,\text{BU},m_n}(t)=\left|\left|\textbf{d}_{q,m_n}^{\text{B,BU}}(t)\right| \right|+\left|\left| \textbf{d}_{p,m_n}^{\text{U,BU}}(t) \right| \right|
\end{equation}
	where $||\cdot||$ means Frobenius norm. $\textbf{d}_{q,m_n}^{\text{B,BI}}(t)$ is the vector pointing from $A_q^{\text{B}}$ to $S_{m_n}^{A,\text{BI}}$. Similarly, $\textbf{d}_{r,m_n}^{\text{I,BI}}(t)$ is the vector from $A_r^{\text{I}}$ to $S_{m_n}^{Z,\text{BI}}$. $\textbf{d}_{r,m_n}^{\text{I,IU}}(t)$ is the vector from $A_r^{\text{I}}$ to $S_{m_n}^{A,\text{IU}}$. $\textbf{d}_{p,m_n}^{\text{U,IU}}(t)$ is the vector from $A_p^{\text{U}}$ to $S_{m_n}^{Z,\text{IU}}$. $\textbf{d}_{q,m_n}^{\text{B,BI}}(t)$ is the vector from $A_q^{\text{B}}$ to $S_{m_n}^{A,\text{BI}}$. $\textbf{d}_{q,m_n}^{\text{B,BU}}(t)$ is the vector from $A_q^{\text{B}}$ to $S_{m_n}^{A,\text{BU}}$. $\textbf{d}_{p,m_n}^{\text{U,BU}}(t)$ is the vector from $A_p^{\text{U}}$ to $S_{m_n}^{Z,\text{BU}}$. $\textbf{d}_{q,m_n}^{\text{B,BI}}(t)$, and $\textbf{d}_{r,m_n}^{\text{I,BI}}(t)$ are calculated as\cite{Bianji}
\begin{equation}
	\textbf{d}_{q,m_n}^{\text{B,BI}}(t)=\textbf{d}_{m_n}^{\text{B,BI}}-[\textbf{l}_{q}^{\text{B}}+\int_{0}^{t}(\textbf{v}^{\text{B}}(t')-\textbf{v}^{A_n}(t'))\,\text{d}t']
\end{equation}
and
\begin{equation}
	\textbf{d}_{r,m_n}^{\text{I,BI}}(t)=\textbf{d}_{m_n}^{\text{I,BI}}-[\textbf{l}_{r}^{\text{I}}+\int_{0}^{t}(-\textbf{v}^{Z_n}(t'))\,\text{d}t']
\end{equation}
	where $\textbf{d}_{m_n}^{\text{B,BI}}$ and $\textbf{d}_{m_n}^{\text{I,BI}}$ are calculated as
\begin{equation}
\begin{split}
	&\textbf{d}_{m_n}^{\text{B,BI}}={d}_{m_n}^{\text{B,BI}}\\
	&[\cos{\phi_{E,m_n}^{\text{B,BI}}}\cos{\phi_{A,m_n}^{\text{B,BI}}},\cos{\phi_{E,m_n}^{\text{B,BI}}}\sin{}\phi_{A,m_n}^{\text{B,BI}},\sin{\phi_{E,m_n}^{\text{B,BI}}}],
\end{split}
\end{equation}
and
\begin{equation}
\begin{split}
	&\textbf{d}_{m_n}^{\text{I,BI}}={d}_{m_n}^{\text{I,BI}}\\
	&[\cos{\phi_{E,m_n}^{\text{I,BI}}}\cos{\phi_{A,m_n}^{\text{I,BI}}},\cos{\phi_{E,m_n}^{\text{I,BI}}}\sin{}\phi_{A,m_n}^{\text{I,BI}},\sin{\phi_{E,m_n}^{\text{I,BI}}}].
\end{split}
\end{equation}
 It should be stressed that all the velocities in this article are 2D vectors. So the elevation angle is zero. And we only give one example of calculating the velocity vector:
\begin{equation}
	\textbf{v}^{\text{B}}(t)=v^\text{B}(t)[\text{cos}\alpha^{\text{B}}(t),\text{sin}\alpha^{\text{B}}(t), 0].
\end{equation}
	In the expression of $\textbf{d}_{q,m_n}^{\text{B,BI}}(t)$, the second item $\textbf{l}_{q}^{\text{B}}$ reflect the spatial non-stationarity and the third item, the integral one shows the time non-stationarity. The other four vectors $\textbf{d}_{r,m_n}^{\text{I,IU}}(t)$, $\textbf{d}_{p,m_n}^{\text{U,IU}}(t)$, $\textbf{d}_{q,m_n}^{\text{B,BU}}(t)$, and $\textbf{d}_{p,m_n}^{\text{U,BU}}(t)$ can be calculated similarly.
	
	In (\ref{channelresponse_BI})-(\ref{channelresponse_BU}), the LoS components are calculated as
\begin{equation}
	h_{qr,\text{BI}}^{L}(t,\tau)=e^{j2\pi f_{c}\tau_{qr,\text{BI}}^{L}(t)}\cdot\delta(\tau-\tau_{qr,\text{BI}}^{L}(t))
\end{equation}
\begin{equation}
	h_{rp,\text{IU}}^{L}(t,\tau)=e^{j2\pi f_{c}\tau_{rp,\text{IU}}^{L}(t)}\cdot\delta(\tau-\tau_{rp,\text{IU}}^{L}(t))
\end{equation}
\begin{equation}
	h_{qp,\text{BU}}^{L}(t,\tau)=e^{j2\pi f_{c}\tau_{qp,\text{BU}}^{L}(t)}\cdot\delta(\tau-\tau_{qp,\text{BU}}^{L}(t))
\end{equation}
	where $\tau_{qr,\text{BI}}^{L}(t)$ is the propogation delay of LoS component between $A_q^{\text{B}}$ and $A_r^{\text{I}}$. $\tau_{rp,\text{IU}}^{L}(t)$ is the propogation delay of LoS component between $A_r^{\text{I}}$ and $A_p^{\text{U}}$. $\tau_{qp,\text{BU}}^{L}(t)$ is the propogation delay of LoS component between $A_q^{\text{B}}$ and $A_p^{\text{U}}$. They are calculated as
\begin{equation}
	\tau_{qr,\text{BI}}^{L}(t)=D_{qr,\text{BI}}(t)/c
\end{equation} 
\begin{equation}
	\tau_{rp,\text{IU}}^{L}(t)=D_{rp,\text{IU}}(t)/c
\end{equation} 
and 
\begin{equation}
	\tau_{qp,\text{BU}}^{L}(t)=D_{qp,\text{BU}}(t)/c
\end{equation} 
	where $D_{qr,\text{BI}}(t)=||\textbf{D}_{qr,\text{BI}}(t)||$ is the distance between $A_q^\text{B}$ and $A_r^\text{I}$ at time instant $t$. The power of the LoS component is set as 1 here on account that the important thing for small scale fading is the relative relationship between power of LoS component and NLoS components which can be adjusted by the Rician factor K. So for the sake of simplicity, we set the power of LoS component as 1. The vector is calculated as \cite{Bianji}
\begin{equation}
	\textbf{D}_{qr,\text{BI}}(t)=\textbf{D}_{\text{BI}}-\textbf{l}_q^{\text{B}}+\textbf{l}_r^{\text{I}}-\int_{0}^{t}\mathbf{v}^{\text{B}}(t')\,\text{d}t'.
\end{equation}
	The other two vectors $\textbf{D}_{rp,\text{IU}}(t)$, and $\textbf{D}_{qp,\text{BU}}(t)$ can be calculated similarly.

	 After obtaining the CIR expressions of three sub-channels, We can apply Fourier transformation to them and obtain transfer functions of different sub-channels, which are $h_{qr,\text{BI}}(t,f)$, $h_{rp,\text{IU}}(t,f)$, and $h_{qp,\text{BU}}(t,f)$. If we omit the steering vector and large scale fading effect, we can obtain the final CIR between $A_{q}^{\text{B}}$ and $A_{p}^{\text{U}}$ which is denoted as $h_{qp,total}(t,f)$. It is calculated as 

\begin{equation}\label{CIRtotal_frequency}
	\begin{split}
	h_{qp,total}(t,f)=&h_{qp,\text{BI}}(t,f)+\\
	&\sum_{r=1}^{M_{xy}}h_{qr,\text{IU}}(t,f)\cdot h_{rp,\text{BU}}(t,f)\cdot e^{j\theta_{r}(t)}
	\end{split}
\end{equation}

	where $\theta_{r}(t)$ is the phase shift that $A_{r}^{\text{I}}$ provides at time instant $t$. According to (\ref{BIU_condition}), it can be calculated as
\begin{equation}
	\theta_{r}(t)=\text{mod}(\frac{2\pi(D_{1r,\text{BI}}(t)+D_{r1,\text{IU}}(t))}{\lambda},2\pi).
\end{equation}

\subsubsection{The Distribution of Scatterers in a Cluster}
	Two coordinates are defined firstly, one is Global Coordinate System (GCS) and the other one is Local Coordinate System (LCS). GCS sets the first antenna element or the center of IRS as its origin. LCS sets the center of the cluster as its origin. $(x,y,z,\phi,\theta)$ is the coordinate value of GCS. The first three values are the coordinate values of the rectangular system. The other two are the angle values of the spherical coordinate system. The value of LCS is $(x',y',z',\phi',\theta')$. We could transform one into the other one through using bearing angle $\alpha$, downtilt angle $\beta$ and slant angle $\gamma$ to generate the rotation matrix \cite{3GPP}. The transformation is based on the theory that any 3D rotation can be divided into three rotations only around rotating axis when the order of rotating axis is determined. Under this condition, the three angles are fixed. The relationship between two sets of rectangular coordinate values is shown as
\begin{equation}
	[x',y',z']=[x,y,z]\cdot\textbf{R}
\end{equation}
	where $R$ is the rotation matrix, which is calculated as
\begin{equation}
\begin{split}
	\textbf{R}=\left[
	\begin{matrix}
	\cos{\alpha}&-\sin{\alpha}&0\\
	\sin{\alpha}&\cos{\alpha}&0\\
	0&0&1
	\end{matrix}
	\right]
	&\left[
	\begin{matrix}
	\cos{\beta}&0&\sin{\beta}\\
	0&1&0\\
	-\sin{\beta}&0&\cos{\beta}
	\end{matrix}
	\right]\\
	\left[
	\begin{matrix}
	1&0&0\\
	0&\cos{\gamma}&-\sin{\gamma}\\
	0&\sin{\gamma}&\cos{\gamma}
	\end{matrix}
	\right].
\end{split}
\end{equation}

	The Gaussian scatterer density model is often used in channel modeling and this assumption is verified by the measurement data \cite{GSDM}. The probability density function of three rectangular values of LCS can be written as \cite{Bianji}
\begin{equation}
	p(x',y',z')=\frac{\text{exp}\left(-\frac{x'^2}{2\sigma_{x}^2}-\frac{y'^2}{2\sigma_{y}^2}-\frac{z'^2}{2\sigma_{z}^2}\right) }{(2\pi)^{\frac{3}{2}}\sigma_x\sigma_y\sigma_z}
\end{equation}
	where $\sigma_{x}$, $\sigma_{y}$, and $\sigma_{z}$ are the variances of three coordinate values, respectively.

\begin{figure}[tb]
	\centerline{\includegraphics[width=0.4\textwidth]{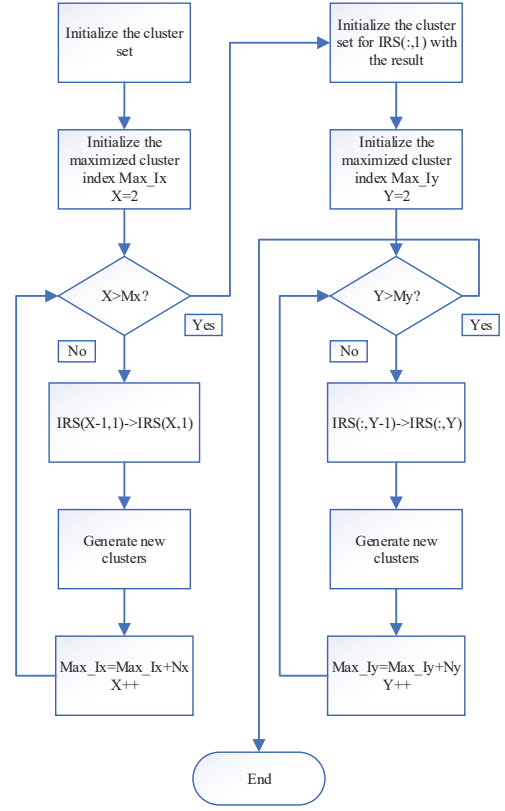}}
	\caption{The evolution procedure of the clusters on the 2D plane.}
	\label{fig_5}
\end{figure}
\subsubsection{Evolution of Clusters in Space Domain}
\begin{figure}[tb]
	\centerline{\includegraphics[width=0.5\textwidth]{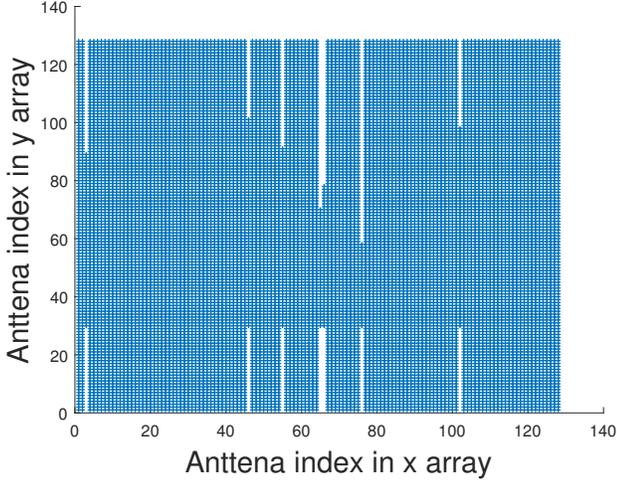}}
	\caption{The evolution of one cluster on a 2D IRS array. ($\lambda_{B}=80$, $\lambda_{D}=4$, $f_c$ = 58 GHz, $M_x=M_y = 128$, $\beta_{E}^{\text{I}_{x}}=\pi/3$, $\beta_{E}^{\text{I}_{y}}=\pi/6$)}
	\label{Cluster_BD}
\end{figure}

	When using the large antenna array, the visibility of the same cluster for different antenna elements is different. This will bring the space non-stationarity. The evolution procedure finishes one direction first and the evolution result is used as the initial state of the second one to evolve on the other direction. The death rate and the generating rate of clusters are denoted as $\lambda_D$ and $\lambda_B$ respectively. The expected value of the initial number of clusters is calculated as $Nc0=\lambda_{B}/\lambda_D$. $\text{IRS}(x,y)$ means the element whose index is $(x,y)$. The death probability on $\text{X}$ direction is calculated as \cite{Bianji}
	\begin{equation} 
		P_{xdeath}=\text{exp}(-\lambda_B\frac{\delta^{I_x}\cos{\beta_E^{I_x}}}{D_C^A}).
	\end{equation}
	where $D_C^A$ is the correlation factor depending on the scenario. The method of cluster evolution from $\text{IRS}(x-1,1)$ to $\text{IRS}(x,1)$ is generating a random vector following the uniform distribution whose number of elements equals to the number of clusters and comparing them with $P_{xdeath}$. If the death probability is larger than the generated element in the random vector, the corresponding cluster is invisible for the next antenna element $\text{IRS}(x,1)$. Then we consider the new cluster generation for $\text{IRS}(x,1)$. The number of new clusters is denoted as $N_x$ which follows the Poisson distribution \cite{3DCM}. The expected value of this variable is calculated as
\begin{equation}
	\text{E}[N_x]=\frac{\lambda_{B}}{\lambda_{D}}(1-P_{xdeath}).
\end{equation}
	The maximum cluster index of the next element increases by this value compared to the last element. For another direction, the different thing is the organization of the uniform distributed variables. The number of the random variables is $M_x \times M_y$. At last, we will obtain a matrix whose dimension is $M_x \times M_y \times Nc0$, containing the information whether the clusters is visible for every antenna element. In Fig. \ref{Cluster_BD}, the simulation result of one cluster evolution on a 2D array for a specific situation is shown. The birth and death rate are set as $\lambda_{B}=80$, and  $\lambda_{D}=4$, respectively. There are $\lambda_{B}/\lambda_{B}=20$ clusters totally. Here we choose the $10\text{th}$ cluster. We can see that its visibility is not the same for elements on IRS array.
	
\begin{figure}[tb]
	\centerline{\includegraphics[width=0.5\textwidth]{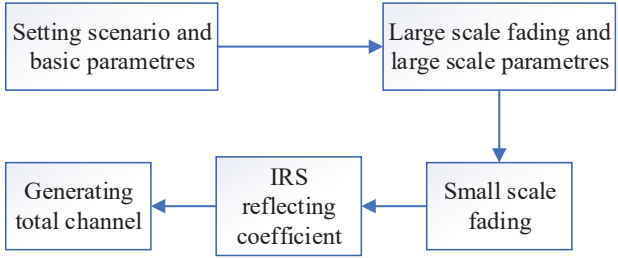}}
	\caption{The whole procedure of generating the channel coefficients matrix.}
	\label{procedure of generating channel matrix}
\end{figure}

	Finally, we can give out the whole procedure of generating the channel coefficients matrix as shown in Fig. \ref{procedure of generating channel matrix}.

\section{Statistical Properties and Results analysis}
	In this section, some typical statistical properties of the proposed non-stationary theoretical IRS channel model are derived. In the following discussion, we will omit the influence of BS-USER sub-channel for the sake of simplicity. That is to say, the cascaded channel assisted by IRS is the main object researched. In fact, in most situations of applying IRS, this sub-channel can be neglected. There are also another two important assumptions, one is that CIR of different sub-channels are uncorrelated, the other one is that LoS component and NLoS component are uncorrelated.
\subsection{Correlation Function}

	Time varying CIR is a kind of random process. Time autocorrelation function (ACF) is defined as the correlation of response value at two time instances. This function reflects how fast channel changes over time. Spatial cross correlation function (CCF) is defined as the correlation of response value between different antenna elements or IRS elements. Similar to the time ACF, this statistical property reflect the rate of change over the antenna array. Time ACF will be discussed first. The situation of one IRS element will be discussed. After that we will talk about the situation with multiple IRS elements and introduce spatial CCF.

	After omitting the BS-USER sub-channel, the transform function between $A_{q}^{\text{B}}$ and $A_{r}^{I}$ can be rewritten as
\begin{equation}\label{CIR_part}
	h_{qp,part}(t,f)=\sum_{r=1}^{M_{xy}}h_{qr,\text{BI}}(t,f)\cdot h_{rp,\text{IU}}(t,f)\cdot e^{j\theta_{r}(t)}.
\end{equation}
	The simulated time varying time ACF between $A_q^\text{B}$ and $A_p^{\text{U}}$ can be shown as
\begin{equation}\label{TACF1}
\begin{split}
		&R_{qp,sim}(t,f;\Delta t)\\
		=&\text{E}\left\{ h_{qp,part}(t,f)h_{qp,part}^{*}(t+\Delta t,f)\right\}  
\end{split}
\end{equation}
	where $(\cdot)^{*}$ denotes the complex conjugate operator. If we consider only one element $A_r^{\text{I}}$ on IRS, equation (\ref{TACF1}) can be rewritten as
\begin{equation}\label{TACF2}
\begin{split}
	&R_{qp,r,sim}(t,f;\Delta t)\\
	=&\text{E}\left\{ h_{qr,\text{BI}}(t,f)h_{qr,\text{BI}}^{*}(t+\Delta t,f)\right\}\times\\
	&\text{E}\left\{ h_{rp,\text{IU}}(t,f)h_{rp,\text{IU}}^{*}(t+\Delta t,f)\right\}\times\\
	&e^{j\theta_{r}(t)-j\theta_{r}(t+\Delta t)}.
\end{split}
\end{equation}
	 The analytical result can be written as
\begin{equation}
\begin{split}
&R_{qp,r,ana}(t,f;\Delta t)\\
=&R_{qr,\text{BI},ana}(t,f;\Delta t)\times R_{rp,\text{IU},ana}(t,f;\Delta t)\\
&\times e^{j\theta_{r}(t)-j\theta_{r}(t+\Delta t)}
\end{split}	
\end{equation}	
	where the analytical result of the time ACF between  $A_q^\text{B}$ and $A_r^\text{I}$ is calculated as
\begin{equation}\label{TACF_BI}
	\begin{split}
		R_{qr,\text{BI},ana}(t,f;\Delta t)=&\frac{K}{K+1}R_{qr,\text{BI}}^{L}(t,f;\Delta t)\\
		&+\frac{1}{K+1}R_{qr,\text{BI}}^{N}(t,f;\Delta t)
	\end{split}
\end{equation}
	where $K$ is the Rician factor. $R_{qr,\text{BI}}^{N}(t,f;\Delta t)$ and $R_{qr,\text{BI}}^{L}(t,f;\Delta t)$ are the NLoS component and LoS component of ACF, which are calculated as
\begin{equation}\label{TACF_BI_NLOS}
\begin{split}
&R_{qr,\text{BI}}^{N}(t,f;\Delta t)\\
=&\sum_{n=1}^{N_{qr,\text{BI}(t)}}\sum_{m_{n}=1}^{M_n}[P_{qr,\text{BI},m_n}(t)P_{qr,\text{BI},m_n}(t+\Delta t)]^{1/2}\\
&e^{j\frac{2\pi(f_c-f)}{c}[d_{qr,\text{BI},m_n}(t)-d_{qr,\text{BI},m_n}(t+\Delta t)]}
\end{split}
\end{equation}
and
\begin{equation}\label{TACF_BI_LOS}
\begin{split}
	R_{qr,\text{BI}}^{L}(t,f;\Delta t)=e^{j\frac{2\pi(f_c-f)}{c}[D_{qr,\text{BI}}(t)-D_{qr,\text{BI}}(t+\Delta t)]}.
\end{split}
\end{equation}
	Similarly, we can obtain the time ACF between $A_{r}^{\text{I}}$ and $A_{p}^{\text{U}}$ as
\begin{equation}\label{TACF_IU}
\begin{split}
		R_{rp,\text{IU},ana}(t,f;\Delta t)=&\frac{K}{K+1}R_{rp,\text{IU}}^{L}(t,f;\Delta t)\\
		&+\frac{1}{K+1}R_{rp,\text{IU}}^{N}(t,f;\Delta t)
\end{split}
\end{equation}
	where the NLoS component and the LoS component can be calculated as
\begin{equation}\label{TACF_IU_NLOS}
\begin{split}
		&R_{rp,\text{IU}}^{N}(t,f;\Delta t)\\
		=&\sum_{n=1}^{N_{rp,\text{IU}(t)}}\sum_{m_{n}=1}^{M_n}[P_{rp,\text{IU},m_n}(t)P_{rp,\text{IU},m_n}(t+\Delta t)]^{1/2}\\
		&e^{j\frac{2\pi(f_c-f)}{c}[d_{rp,\text{IU},m_n}(t)-d_{rp,\text{IU},m_n}(t+\Delta t)]}
\end{split}
\end{equation}
and
\begin{equation}\label{TACF_IU_LOS}
\begin{split}
		R_{rp,\text{IU}}^{L}(t,f;\Delta t)=e^{j\frac{2\pi(f_c-f)}{c}[D_{rp,\text{IU}}(t)-D_{rp,\text{IU}}(t+\Delta t)]}.
\end{split}
\end{equation}
	Then we will talk about the situation considering all the elements on IRS. After substituting (\ref{CIR_part}) into (\ref{TACF2}), we can obtain 
\begin{equation}\label{TACF_r1r2}
\begin{aligned}
	&R_{qp,sim}(t,f;\Delta t)\\
	=&\text{E}\left\lbrace{\sum_{r=1}^{M_{xy}}h_{qr,\text{BI}}(t,f)h_{rp,\text{IU}}(t,f)e^{j\theta_{r}(t)}}\right. \\
	&\phantom{=\;\;}\left. \sum_{r=1}^{M_{xy}}h_{qr,\text{BI}}^{*}(t+\Delta t,f)h_{rp,\text{IU}}^{*}(t+\Delta t)e^{-j\theta_{r}(t+\Delta t)}\right\rbrace \\
	=&\text{E}\left\lbrace \sum_{r_{1}=1}^{M_{xy}}\sum_{r_{2}=1}^{M_{xy}}h_{qr_{1},\text{BI}}(t,f)h_{r_{1}p,\text{IU}}(t,f)\right.\\
	&\phantom{=\;\;}\left.h_{qr_{2},\text{BI}}^{*}(t+\Delta t,f)h_{r_{2}p,\text{IU}}^{*}(t+\Delta t) e^{j\theta_{r_{1}}(t)-j\theta{r_{2}}(t+\Delta t)}\right\rbrace\\
	=&\sum_{r_{1}=1}^{M_{xy}}\sum_{r_{2}=1}^{M_{xy}}\text{E}\left\lbrace h_{qr_{1},\text{BI}}(t,f)h_{qr_{2},\text{BI}}^{*}(t+\Delta t,f) \right\rbrace\\
	&\text{E}\left\lbrace h_{r_{1}p,\text{IU}}(t,f)h_{r_{2}p,\text{IU}}^{*}(t+\Delta t)\right\rbrace e^{j\theta_{r_{1}}(t)-j\theta_{r_{2}}(t+\Delta t)}
\end{aligned}
\end{equation}
	Let 
\begin{equation}
	R_{qr_{1},qr_{2},\text{BI}}(t,f;\Delta t)=\text{E}\left\lbrace h_{qr_{1},\text{BI}}(t,f)h_{qr_{2},\text{BI}}^{*}(t+\Delta t,f) \right\rbrace
\end{equation}
and
\begin{equation}
	R_{r_{1}p,r_{2}p,\text{IU}}(t,f;\Delta t)=\text{E}\left\lbrace h_{r_{1}p,\text{IU}}(t,f)h_{r_{2}p,\text{IU}}^{*}(t+\Delta t)\right\rbrace
\end{equation}
	where $R_{qr_{1},qr_{2},\text{BI}}(t,f;\Delta t)$ and $R_{r_{1}p,r_{2}p,\text{IU}}(t,f;\Delta t)$ are the spatial CCF of BS-IRS sub-channel and IRS-USER sub-channel. Equation (\ref{TACF_r1r2}) can be rewritten as 
\begin{equation}
\begin{aligned}
	&R_{qp,sim}(t,f;\Delta t)\\
	=&\sum_{r_{1}=1}^{M_{xy}}\sum_{r_{2}=1}^{M_{xy}}R_{qr_{1},qr_{2},\text{BI}}(t,f;\Delta t)R_{r_{1}p,r_{2}p,\text{IU}}(t,f;\Delta t)\\
	&e^{j\theta_{r_{1}}(t)-j\theta_{r_{2}}(t+\Delta t)}.
\end{aligned}
\end{equation}
   Spatial CCF can also be expressed as the sum of LoS component and NLoS component. $R_{qr_{1},qr_{2},\text{BI}}(t,f;\Delta t)$ can be calculated as
\begin{equation}
	\begin{split}
	R_{qr_{1},qr_{2},\text{BI}}(t,f;\Delta t)=&\frac{K}{K+1}R_{qr_{1},qr_{2},\text{BI}}^{L}(t,f;\Delta t)\\
	&+\frac{1}{K+1}R_{qr_{1},qr_{2},\text{BI}}^{N}(t,f;\Delta t)
	\end{split}
\end{equation}
	where $R_{qr_{1},qr_{2},\text{BI}}^{N}(t,f;\Delta t)$ is calculated as
\begin{equation}
\begin{split}
	&R_{qr_{1},qr_{2},\text{BI}}^{N}(t,f;\Delta t)\\
	=&\sum_{n=1}^{N_{qr_{1},qr_{2},\text{BI}(t)}}\sum_{m_{n}=1}^{M_n}[P_{qr_{1},\text{BI},m_n}(t)P_{qr_{2},\text{BI},m_n}(t+\Delta t)]^{1/2}\\
	&e^{j\frac{2\pi(f_c-f)}{c}[d_{qr_{1},\text{BI},m_n}(t)-d_{qr_{2},\text{BI},m_n}(t+\Delta t)]}
\end{split}
\end{equation}
	where $N_{qr_{1},qr_{2},\text{BI}}(t)$ is the smaller value of $N_{qr_{1},\text{BI}}(t)$ and $N_{qr_{2},\text{BI}}(t)$. In fact, the numbers of clusters between $A_{q}^{\text{B}}$ and $A_{r_{1}}^{\text{I}}$ and between $A_{q}^{\text{B}}$ and $A_{r_{2}}^{\text{I}}$ are usually set as the same value when we simulate the correlation function because in practice the difference is very small. Similar to (\ref{TACF_BI_LOS}), $R_{qr_{1},qr_{2},\text{BI}}^{L}(t,f;\Delta t)$ is calculated as
\begin{equation}
	\begin{split}
	&R_{qr_{1},qr_{2},\text{BI}}^{L}(t,f;\Delta t)\\
	=&e^{j\frac{2\pi(f_c-f)}{c}[D_{qr_{1},\text{BI}}(t)-D_{qr_{2},\text{BI}}(t+\Delta t)]}.
	\end{split}
\end{equation}
	$R_{r_{1}p,r_{2}p,\text{IU}}(t,f;\Delta t)$ can be obtained through the same way.
	
\subsection{Root Mean Square (RMS) Delay Spread Cumulative Distribution Function (CDF)}
	Delay spread reflect the largest propogation delay difference among all the rays received by USER. Here we calculate this physical quantity statistically, which means RMS delay spread. To get the CDF of RMS delay spread, we should operate the channel simulation many times and count these results then we can obtain the RMS delay spread CDF. The RMS delay spread of the sub-channel from $A_{q}^{\text{B}}$ to $A_{r}^{\text{I}}$ can be calculated as
\begin{equation}
\begin{aligned}
	&DS_{qr,\text{BI}}(t)=\left( \sum_{n=1}^{N_{qr,\text{BI}}(t)}\sum_{m_n=1}^{M_n}P_{qr,\text{BI},m_n}(\tau_{qr,\text{BI},m_n})^2\right. \\
	&\phantom{=\;\;}\left. -(\sum_{n=1}^{N_{qr,\text{BI}}(t)}\sum_{m_n=1}^{M_n}P_{qr,\text{BI},m_n}\tau_{qr,\text{BI},m_n})^2\right)^{\frac{1}{2}}.
\end{aligned}
\end{equation}

\subsection{Local Doppler Spread}
	The instantaneous frequency is important for signal recognition, estimation, and modeling on account that it can provide a energy distribution on frequency domain. Here we use the Doppler frequency to represent the instantaneous frequency. It is given by \cite{Bianji}
\begin{equation}
\begin{split}
	\nu_{qr,m_n}(t)=&\frac{1}{\lambda}\frac{\text{d}[d_{q,m_n}^{\text{B,BI}}(t)+d_{r,m_n}^{\text{I,BI}}(t)]}{dt}\\
	&+\frac{1}{\lambda}\frac{\text{d}[d_{r,m_n}^{\text{I,IU}}(t)+d_{p,m_n}^{\text{U,IU}}(t)]}{dt}.
\end{split}
\end{equation}
	Note that Doppler frequency changes with time because of the motion of the clusters and the communication terminals. At last, the local Doppler spread can be calculated as \cite{Bianji}
\begin{equation}
	B_{qr}(t)=(\text{E}[\nu_{qr,m_n}(t)^2]-\text{E}[\nu_{qr,m_n}(t)]^2)^{\frac{1}{2}}.
\end{equation}
\section{Results and Analysis}
\subsection{Time ACF}
\begin{figure}[tb]
	\centerline{\includegraphics[width=0.5\textwidth]{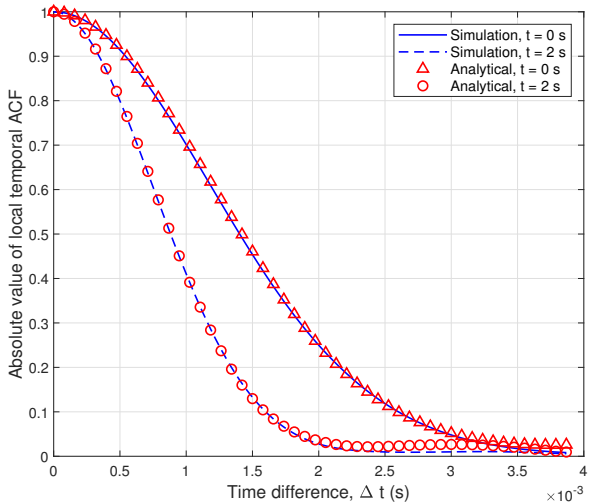}}
	\caption{The comparison of time ACF between the simulation and the analytical result at $t$ = 0 s and $t$ = 2 s. ($D_{\text{BI}}$ = 100 m, $D_{\text{IU}}$ = 200 m, $f_c$ = 62 GHz, $v^{\text{B}}$ = 10 m/s, $v^{\text{U}}$ = 10 m/s, $q$ = 1, $p$ = 1, $r$ = 1)}
	\label{timeACF1}
\end{figure}
\begin{figure}[tb]
	\centerline{\includegraphics[width=0.5\textwidth]{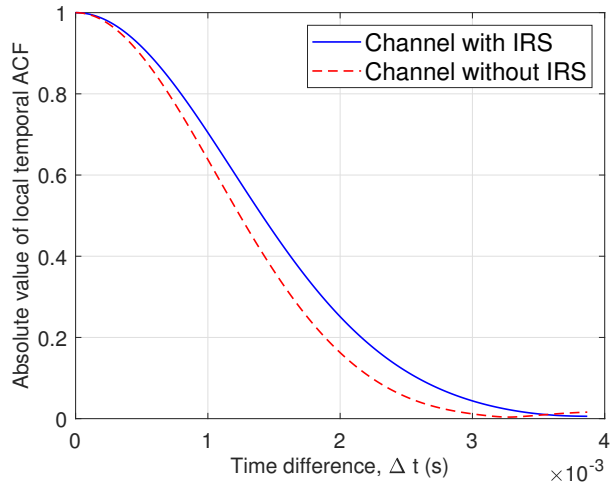}}
	\caption{The comparison of time ACF between the channel using IRS with one element and without IRS. ($v^{\text{B}}$ = 10 m/s, $v^{\text{U}}$ = 10 m/s, $D_{\text{BI}}$ = 100 m, $f_c$ = 62 GHz)}
	\label{timeACF comparison}
\end{figure}
\begin{figure}[tb]
	\centerline{\includegraphics[width=0.5\textwidth]{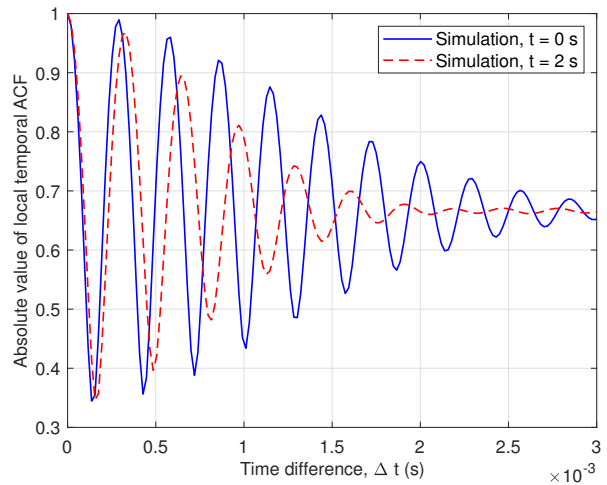}}
	\caption{The comparison of time ACF when existing a LoS component. ($K$ = 5 dB, the other parameters are the same with Fig. \ref{timeACF1}.)}
	\label{LOSTACF}
\end{figure}
\begin{figure}[tb]
	\centerline{\includegraphics[width=0.5\textwidth]{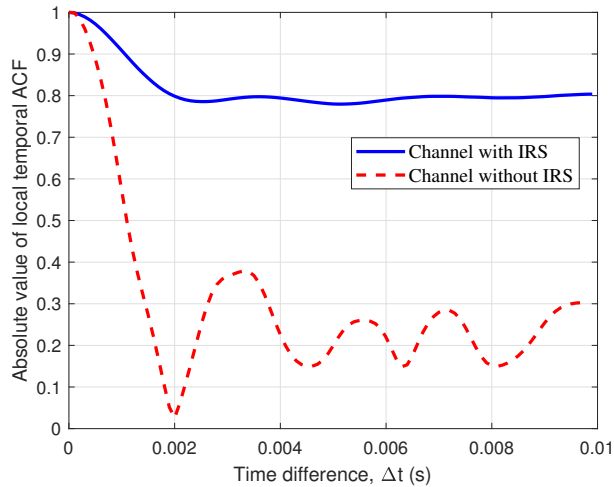}}
	\caption{The comparison of time ACF between the situations with and without IRS . ($M_{\text{B}}$ = 1, $M_{\text{U}}$ = 1, $M_{x}$ =$M_{y}$ = 2,  $f_c$ = 10 GHz, $K$ = 0.02 dB, $D_{\text{BU}}$ = 150 m, $D_{\text{BI}}$ = 50 m, $v^{\text{B}}$ = 0 m/s, $v^{\text{U}}$ = 10 m/s, $t$ = 2 s )}
	\label{sumTACF_noIRS}
\end{figure}
\begin{figure}[tb]
	\centerline{\includegraphics[width=0.5\textwidth]{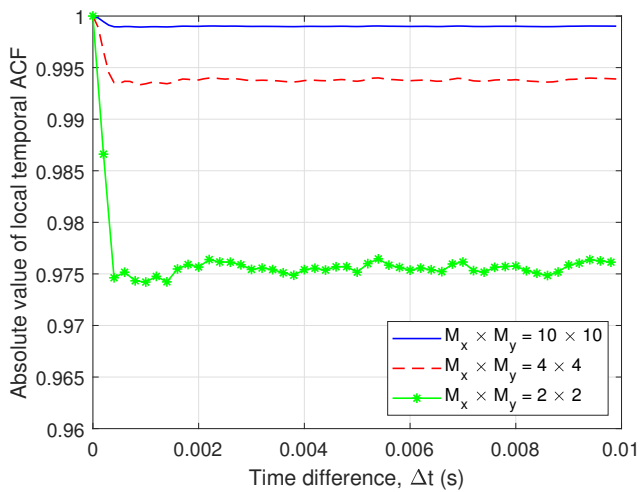}}
	\caption{The comparison of time ACF when using IRS with different sizes. ($M_{\text{B}}$ = 1, $M_{\text{U}}$ = 1, $f_c$ = 58 GHz, $K$ = 5 dB, $D_{\text{BU}}$ = 200 m, $D_{\text{BI}}$ = 50 m, $v^{\text{B}}$ = 10 m/s, $v^{\text{U}}$ = 10 m/s, $t$ = 2 s)}
	\label{sumTACF_scale}
\end{figure}
\begin{figure}[tb]
	\centerline{\includegraphics[width=0.5\textwidth]{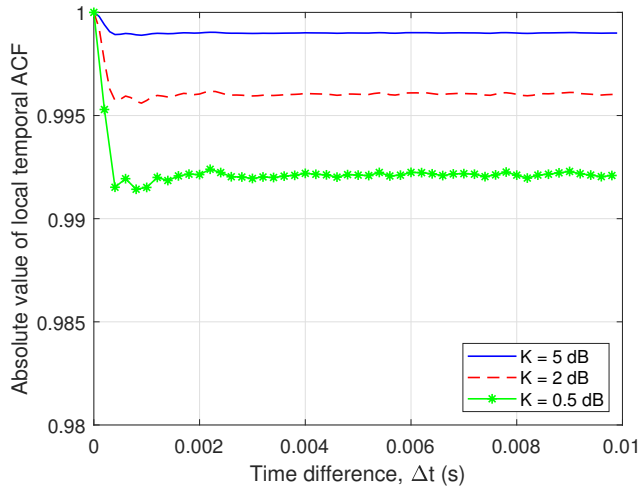}}
	\caption{The comparison of time ACF with different Rician factor. ($M_{\text{B}}$ = 1, $M_{\text{U}}$ = 1, $M_{x}$ =$M_{y}$ = 10,  $f_c$ = 58 GHz, $D_{\text{BU}}$ = 200 m, $D_{\text{BI}}$ = 50 m, $v^{\text{B}}$ = 10 m/s, $v^{\text{U}}$ = 10 m/s, $t$ = 2 s )}
	\label{sumTACF_Rice}
\end{figure}
\begin{figure}[tb]
	\centerline{\includegraphics[width=0.5\textwidth]{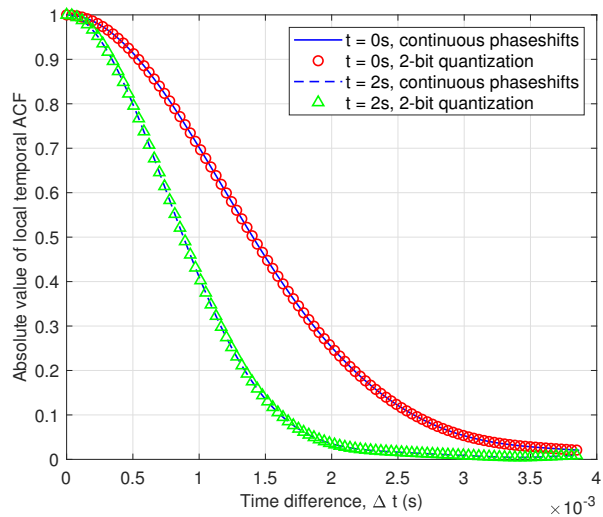}}
	\caption{The comparison of time ACF when using one element IRS with practical phase shifts. (The parameters setting is the same with Fig. \ref{timeACF1})}
	\label{TACFQ}
\end{figure}
\begin{figure}[tb]
	\centerline{\includegraphics[width=0.5\textwidth]{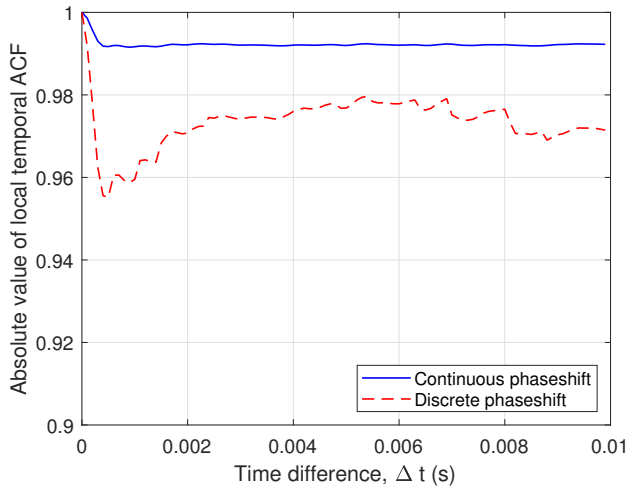}}
	\caption{The comparison of time ACF when using IRS with 2-bit phase shifts considering all the elements. ($M_{\text{B}}$ = 1, $M_{\text{U}}$ = 1, $f_c$ = 58 GHz, $K$ = 5 dB, $D_{\text{BU}}$ = 200 m, $D_{\text{BI}}$ = 50 m, $v^{\text{B}}$ = 10 m/s, $v^{\text{U}}$ = 10 m/s, $t$ = 2 s)}
	\label{discreteTACF}
\end{figure}

	 Fig \ref{timeACF1} shows time ACF at different time instances considering only one element of IRS. The agreement between simulation result and analytical result verifies the correctness of the proposed channel model. The disagreement of time ACF at two time instants confirms the time non-stationarity of the proposed channel model. Fig. \ref{timeACF comparison} considers only one element on IRS. It shows that the value of time ACF is higher after we use IRS. The main reason causing this result under one element situation is that IRS plays a role in separating the whole channel into two parts, making the expression of time ACF of the cascaded channel the product of two sub-channels' ACF as shown in (\ref{TACF2}). It should be noted that we do not consider the LoS component both in Fig. \ref{timeACF1} and Fig. \ref{timeACF comparison}. The simulation result considering LoS component is shown in Fig. \ref{LOSTACF}. We can see that the fluctuation and the value of ACF are both much huger than before. This is also because we consider the time ACF of only one IRS element. Only one LoS component between a pair of antenna elements can be adjusted by IRS. If all the elements on IRS and LoS components are considered, ACF is larger and tends to be flat with time interval increasing as shown in Fig. \ref{sumTACF_noIRS}. We can also see that larger size of IRS causes larger value of time ACF. The simulation result is shown in Fig. \ref{sumTACF_scale}. Another interesting phenomenon is that time correlation function value is also affected by the Rician factor. This conclusion is proved by the simulation result shown in Fig. \ref{sumTACF_Rice}. We can find that the stronger the LoS component is and the larger the size of IRS is, the larger ACF will be. This is because IRS can make full use of the LoS components and eliminate the phase differences among different LoS components through providing extra phase shifts. So Rician factor and size can affect the channel characteristics strongly. The larger the size of IRS is, the more LoS components IRS can adjust and the stronger the correlation among CIR of different time instances will be. IRS makes the ACF value raised. In combination with the definition of correlation function we give out before, we can draw a conclusion that IRS makes the stability of channel stronger, which is good for information transportation. As shown in Fig. \ref{TACFQ}, practical phase shifts do not affect the time ACF with respect to the continuous phase shifts if there is only one element on IRS. This is because IRS introduces two phase product factors into ACF, not affecting the absolute value of ACF. So we can not find the differences when observing the absolute value of this function. But as shown in Fig. \ref{discreteTACF}, when we consider all the elements on IRS, the resolution of IRS will affect the value of ACF. This is because ACF is the sum of multiple complex numbers. The phase difference among these numbers will influence the result of adding them together.

\subsection{Spatial CCF}
\begin{figure}[tb]
	\centerline{\includegraphics[width=0.5\textwidth]{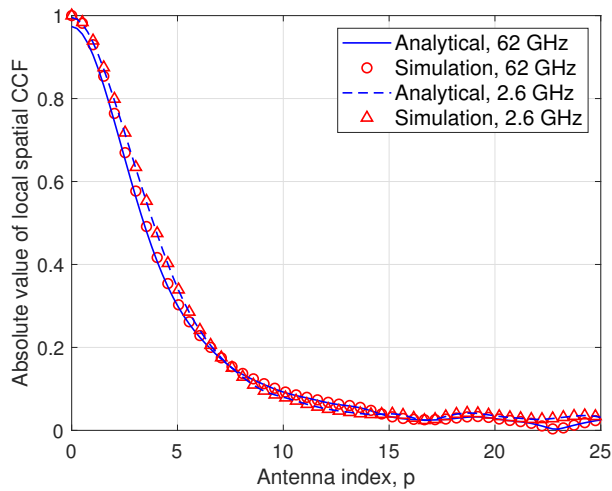}}
	\caption{The comparison of spatial CCF between the simulation and the analytical result at 62 GHz and 2.6 GHz.($M_{\text{B}}$ = 100, $M_{\text{U}}$ = 1, $D_{\text{BU}}$ = 100 m, $v^{\text{B}}$ = 10 m/s, $v^{\text{U}}$= 10 m/s)}
	\label{spaceCCF}
\end{figure}

	As shown in Fig. \ref{spaceCCF}, we can see that two curves are almost the same because the antenna interval is small due to the high frequency making the difference very small. At the same time, the fact that analytical results and simulation results match well indicates the accuracy and the correctness of the proposed channel model.

\subsection{Local Doppler Spread}
\begin{figure}[tb]
	\centerline{\includegraphics[width=0.5\textwidth]{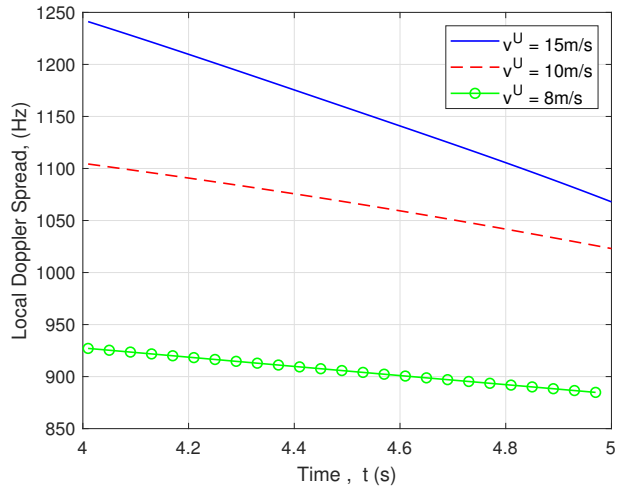}}
	\caption{The comparison of spatial CCF with different velocities of USER.($D_{\text{BI}}$ = 100 m,$D_{\text{IU}}$ = 200 m, $v^{\text{B}}$ = 0 m/s, $v^{Z_n}$ = 5 m/s, $v^{\text{U}}$ = 15 m/s or 10 m/s or 8 m/s)}
	\label{doppler}
\end{figure}

	Under the conditions of different velocities of USER, we can see that higher velocity results in higher local Doppler spread in Fig. \ref{doppler}. When we focus on one curve with a fixed speed of USER, we can find that local Doppler spread is decreasing with time. It changes with time instants just because of the motions of the clusters and USER at the same time in different directions.
\subsection{RMS Delay Spread CDF}
\begin{figure}[tb]
	\centerline{\includegraphics[width=0.5\textwidth]{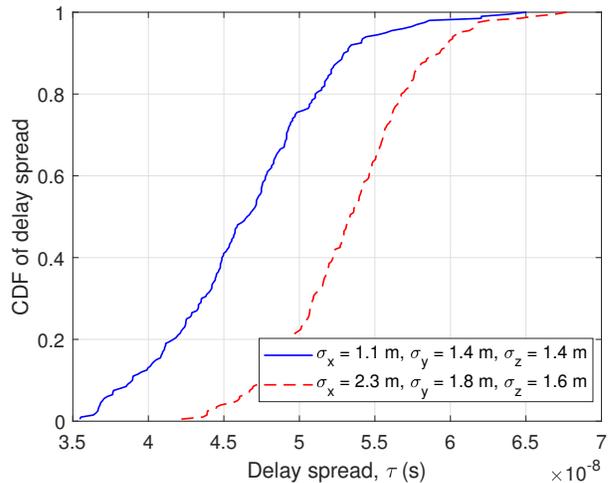}}
	\caption{The comparison of RMS delay spread CDF between the different variance of the scatterers' coordinate values.}
	\label{DS_CDF}
\end{figure}
	As shown in Fig. \ref{DS_CDF}, we can see that under the condition of more dispersed scatterers, the delay spread is bigger. This is because the signal that travels the longest distance among all the rays needs to travel a longer distance in the environment with more dispersed scatterers.
\section{Conclusions}
In this paper, a GBSM for IRS-based 6G channel has been proposed. The evolution of time and space have been considered through the time varying distance and cluster evolution matrix. The statistical properties such as time ACF, spatial CCF, local Doppler spread, and DS CDF have been simulated and analyzed. The good agreement illustrates the correctness of the proposed channel model. The differences among the curves demonstrate the non-stationary properties of the proposed channel model. The fact that time ACF's value of the situation using IRS is higher than it without IRS illustrates that IRS can separate the channel and change the statistical properties of the channel. The comparison between discrete and continuous phase shifts help us draw a conclusion that discrete phase shifts do not change the absolute value of time ACF.


	\begin{IEEEbiography}[{\includegraphics[width=1in,clip,keepaspectratio]{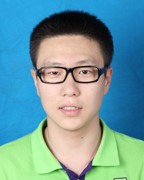}}]{Yingzhuo Sun}
	received the B.E. degree in Information Engineering from Southeast University, China, in 2020. He is currently pursuing the M.Sc. degree in the National Mobile
	Communications Research Laboratory, Southeast University, China. His research interests are IRS wireless channel measurements and modeling.
\end{IEEEbiography}

\begin{IEEEbiography}[{\includegraphics[width=1in,height=1.25in,clip,keepaspectratio]{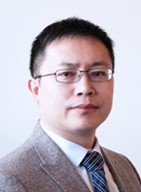}}]{Cheng-Xiang Wang}
(S'01-M'05-SM'08-F'17) received the BSc and MEng degrees in
Communication and Information Systems from Shandong University, China, in 1997 and 2000, respectively, and the PhD degree in Wireless Communications from Aalborg University, Denmark, in 2004.

He was a Research Assistant with the Hamburg University of Technology, Hamburg, Germany, from 2000 to 2001, a Visiting Researcher with
Siemens AG Mobile Phones, Munich, Germany, in 2004, and a Research
Fellow with the University of Agder, Grimstad, Norway, from 2001 to 2005. He has been with Heriot-Watt University, Edinburgh, U.K., since 2005, where he was promoted to a Professor in 2011. In 2018, he joined the National Mobile Communications Research Laboratory, Southeast University, China, as a Professor. He is also a part-time professor with the Purple Mountain Laboratories, Nanjing, China. He has authored four books, three book chapters, and more than 410 papers in refereed journals and conference proceedings, including 24 Highly Cited Papers. He has also delivered 22 Invited Keynote Speeches/Talks and~7 Tutorials in international conferences. His current research interests include wireless channel measurements and modeling, 6G wireless communication networks, and applying artificial intelligence to wireless networks.
	
Prof. Wang is a member of the Academia Europaea, a fellow of the IET, an IEEE Communications Society Distinguished Lecturer in 2019 and 2020, and a Highly-Cited Researcher recognized by Clarivate Analytics in 2017-2020. He is currently an Executive Editorial Committee member for the IEEE TRANSACTIONS ON WIRELESS COMMUNICATIONS. He has served as an Editor for nine international journals, including the IEEE TRANSACTIONS ON WIRELESS
COMMUNICATIONS from 2007 to 2009, the IEEE TRANSACTIONS
ON VEHICULAR TECHNOLOGY from 2011 to 2017, and the
IEEE TRANSACTIONS ON COMMUNICATIONS from 2015 to 2017.
He was a Guest Editor for the IEEE JOURNAL ON SELECTED AREAS
IN COMMUNICATIONS, Special Issue on Vehicular Communications and
Networks (Lead Guest Editor), Special Issue on Spectrum and Energy
Efficient Design of Wireless Communication Networks, and Special Issue
on Airborne Communication Networks. He was also a Guest Editor for  the IEEE TRANSACTIONS ON BIG DATA, Special Issue on Wireless  Big Data, and is a Guest Editor for the IEEE TRANSACTIONS ON  COGNITIVE COMMUNICATIONS AND NETWORKING, Special Issue on Intelligent Resource Management for 5G and Beyond. He has served as a TPC Member, a TPC Chair, and a General Chair for more than 80 international conferences. He received 12 Best Paper Awards from IEEE GLOBECOM 2010, IEEE ICCT 2011, ITST 2012, IEEE VTC 2013-Spring, IWCMC 2015, IWCMC 2016, IEEE/CIC ICCC 2016, WPMC 2016, WOCC 2019, IWCMC 2020, and WCSP 2020. 
\end{IEEEbiography}

\begin{IEEEbiography}[{\includegraphics[width=1in,height=1.25in,clip,keepaspectratio]{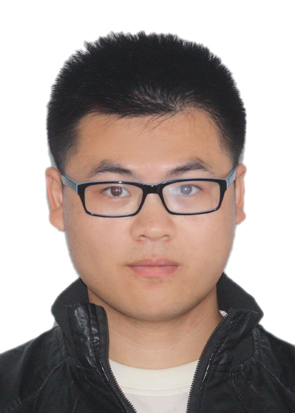}}]{Jie Huang}
	(M’20) received the B.E. degree in Information Engineering from Xidian University, China, in 2013, and the Ph.D. degree in Communication and Information Systems from Shandong University, China, in 2018. From October 2018 to October 2020, he was a Postdoctoral Research Associate in the National Mobile Communications Research Laboratory, Southeast University, China. From January 2019 to February 2020, he was a Postdoctoral Research Associate in Durham University, UK. He is currently an Associate Professor·in Southeast University and a researcher in Purple Mountain Laboratories, China. His research interests include millimeter wave, THz, massive MIMO, intelligent reflecting surface channel measurements and modeling, wireless big data, and 6G wireless communications.
\end{IEEEbiography}

\begin{IEEEbiography}[{\includegraphics[width=1in,clip,keepaspectratio]{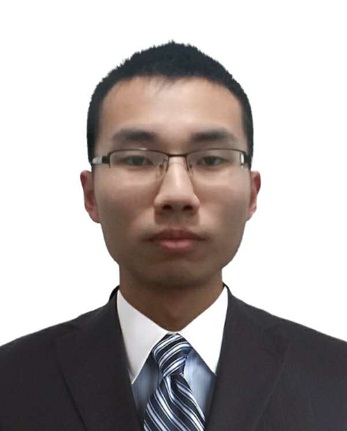}}]{Jun Wang}
	received the B.E. degree in Information Engineering from Southeast University, China, in 2016. He is currently pursuing the Ph.D. degree in the National Mobile
	Communications Research Laboratory, Southeast University, China. His research interests are THz wireless channel measurements and modeling.
\end{IEEEbiography}

\end{document}